\documentclass[a4paper,reqno]{amsart}
\usepackage{fullpage}
\usepackage[latin1]{inputenc}
\usepackage[T1]{fontenc}
\usepackage{amsfonts}
\usepackage{color}
\usepackage{epsfig}
\usepackage{epstopdf}
\usepackage{subfigure}
\usepackage{bm}
\usepackage{bbm}
\usepackage{bbold}
\usepackage{amssymb}
\usepackage{amsmath}
\usepackage{amsthm}
\usepackage[foot]{amsaddr}
\usepackage{graphicx}
\usepackage{subfigure}
\usepackage[dvipsnames]{xcolor}
\usepackage{afterpage}
\usepackage{booktabs}
\usepackage{siunitx}

\usepackage{booktabs}
\usepackage{fix-cm}

\usepackage[margin=2.5cm]{geometry}

\usepackage[colorlinks=true, linkcolor=blue, citecolor=cyan, urlcolor=blue]{hyperref}

\usepackage{array}
\usepackage{makecell}

\usepackage[todonotes={textsize=footnotesize}]{changes}
\definechangesauthor[color=red]{FL}
\definechangesauthor[color=blue]{SZ}
\definechangesauthor[color=orange]{CC}
\definechangesauthor[color=green]{PM}

\theoremstyle{definition}
\newtheorem{definition}{Definition}[section]

\numberwithin{equation}{section}

\definecolor{light}{gray}{.9}

\def\be{\begin{equation}}
\def\ee{\end{equation}}
\def\bea{\begin{eqnarray}}
\def\eea{\end{eqnarray}}

\def\E{\mathbb{E}}

\def\X{\bm{X}}
\def\Y{\bm{Y}}

\def\diag{\operatorname{diag}}

\newcommand{\ie}{\textit{i.e. }}
\newcommand{\eg}{\textit{e.g. }}

\newcommand{\nocontentsline}[3]{}
\newcommand{\tocless}[2]{\bgroup\let\addcontentsline=\nocontentsline#1{#2}\egroup}

\DeclareMathSymbol{\leqslant}{\mathalpha}{AMSa}{"36} % nicer `smaller or equal'
\DeclareMathSymbol{\geqslant}{\mathalpha}{AMSa}{"3E} % nicer `larger or equal'
\DeclareMathSymbol{\eset}{\mathalpha}{AMSb}{"3F}     % nicer `emptyset'
\renewcommand{\leq}{\;\leqslant\;}                   % redef. of < or =
\renewcommand{\geq}{\;\geqslant\;}                   % redef. of > or =

\def\P{\mathbb{P}}
\def\E{\mathbb{E}}

\title{%Tail Granger Causalities and where to find them:\\ true interactions and spurious effects
Tail Granger causalities and where to find them:\\
extreme risk spillovers vs spurious linkages}
\date{October 24, 2020}

\author[1]{Piero Mazzarisi\textnormal{$^{1,2,*}$}}

\author[2]{Silvia Zaoli$^{1,3}$}

\author[3]{Carlo Campajola$^{2}$}

\author[4]{Fabrizio Lillo$^{1}$}

\address{$^{(1)}$University of Bologna, Department of Mathematics}
\address{$^{(2)}$Scuola Normale Superiore, Pisa}
\address{$^{(3)}$Quantitative Life Science Section, The Abdus Salam International center for Theoretical  Physics (ICTP), Trieste}
\address{\textnormal{$^*$Corresponding author at: Scuola Normale Superiore, Piazza dei Cavalieri, 7, 56126 Pisa (PI), Italy}}

\email{piero.mazzarisi@sns.it, szaoli@ictp.it, carlo.campajola@sns.it, fabrizio.lillo@unibo.it}


\begin{document}
\maketitle
\begin{abstract}
%Indicators of extreme downside market risk are of great interest to control financial risk and for portfolio management. Granger causality in tail can be thus defined to capture spillover effects between extreme price movements of financial instruments. Here, in the same spirit of the original test by Granger, we introduce a novel test of Granger causality in tail which adopts the likelihood ratio statistic and is based on the vector generalization of the {\it discrete autoregressive process} for binary time series describing the sequence of extreme events of some underlying price dynamics. The proposed test has very good size and power in finite samples, especially for large sample size, it allows inferring the correct time scale at which the causal interaction takes place, and it is flexible enough for multivariate extension when more than two time series are considered in order to decrease false detections as spurious effect of neglected variables. A simulation study shows the performances of the proposed method with a large variety of data generating processes and it introduces also the comparison with the test of Granger causality in tail by \cite{hongetal}. We report both advantages and drawbacks of the different approaches, thus pointing out some crucial aspects related to the false detections of Granger causality for tail events. An empirical application to high frequency data of the US stock exchange market highlights the merit of our novel approach.
Identifying risk spillovers in financial markets is of great importance for assessing systemic risk and portfolio management. Granger causality in tail (or in risk) tests whether past extreme events of a time series help predicting future extreme events of another time series. The topology and connectedness of networks built with Granger causality in tail can be used to measure systemic risk and to identify risk transmitters. Here we introduce a novel test of Granger causality in tail which adopts the likelihood ratio statistic and is based on the multivariate generalization of a discrete autoregressive process for binary time series describing the sequence of extreme events of the underlying price dynamics. The proposed test has very good size and power in finite samples, especially for large sample size, allows inferring the correct time scale at which the causal interaction takes place, and it is flexible enough for multivariate extension when more than two time series are considered in order to decrease false detections as spurious effect of neglected variables. An extensive simulation study shows the performances of the proposed method with a large variety of data generating processes and it introduces also the comparison with the test of Granger causality in tail by \cite{hongetal}. We report both advantages and drawbacks of the different approaches, pointing out some crucial aspects related to the false detections of Granger causality for tail events. An empirical application to high frequency data of a portfolio of US stocks highlights the merits of our novel approach.
\end{abstract}
%\tableofcontents

\section*{Introduction}

The problem of inferring causal interactions from data has a remarkable history in scientific research and the milestone work of Granger \cite{granger} represented the turning point in such study. According to Granger causality, given a couple of time series $x$ and $y$, it is said that $y$ `Granger-causes' $x$ if the past information on $y$ helps in forecasting $x$ better than using only the past information on $x$. Granger causality overcomes the philosophical question of properly defining what `true causality' means, by limiting the study to systems whose state can be assessed quantitatively by means of time series and relying on the concept of `predictive causality' \cite{granger2}. Within this framework, Granger causality is pragmatic, well defined, and has exhibited many successful applications in a variety of fields, from quantitative finance \cite{billio,pirino} to transportations \cite{zanin} and neuroscience \cite{seth}.

In time series econometrics, the most commonly used test of Granger causality for bivariate systems is the F-test originally proposed in \cite{granger}. This test is sometimes referred to as causality {\it in mean} since the statistical testing procedure is based on the evaluation of the forecasting performances associated with the first moment of a time series. The most stringent assumption consists in considering the information on the two time series as all the {\it significant} information when testing for Granger causality. This is a strong assumption because, in the case of a high dimensional system, a low dimensional subprocess contains little information about the true structure of interactions and some causal relations might be falsely detected as spurious effect of neglected variables \cite{helmut}. Starting with the seminal work of \cite{geweke}, multivariate approaches have been proposed to correct these spurious effects by taking into account network effects in the statistical testing procedure. Finally, Granger causality has been investigated from a theoretical point of view moving from Econometrics to Information Theory and, in particular, the equivalence with transfer entropy was proved in the Gaussian case \cite{barnett1}, which is in turn equivalent to the log-likelihood ratio statistic \cite{barnett2}. In practice, this implies, among other things, that the Likelihood-Ratio test is supported as statistical method for inferring Granger causality. The advantage relies on less stringent hypotheses about the statistical distributions with respect to the F-test.

Granger causality proved useful in monitoring systemic risk in financial markets. In fact, recent applications showed that it captures the network of risk propagation between market participants, and the degree of interconnectedness of this network can be used to define indicators of systemic risk. For instance, \cite{billio} have adopted Granger causality {\it in mean} as a proxy for return-spillover effects among hedge funds, banks, broker/dealers, and insurance companies (\eg see \cite{danielsson,battiston,duarte} for theoretical studies on spillover effects during financial crises), showing that the financial system has become considerably more interconnected before the financial crisis of 2007-2008 because financial innovations and deregulation had increased the interdependence of business between such investors. 

The original Granger causality test evaluates the forecasting performance giving equal importance to the ability to forecast average or extreme values, negative or positive ones. However, when monitoring financial risk, extreme downside market movements are much more important than small fluctuations for spillover effects. A  method specific for risk measures, in particular volatility, was introduced by \cite{cheung} with the concept of Granger causality {\it in variance}, by extending the concept of causation to the second moment. Nevertheless, variance  is a two-sided risk measure and it is not able to capture heavy tails, thus the causal relations between extreme events of two time series. To this end, Granger causality {\it in tail} can be defined. The concept was firstly introduced by \cite{hongetal}. In this work, the authors have proposed a kernel-based test to detect extreme downside risk spillovers with a statistical procedure in two steps: (i) measuring risk by the left (or right, depending on the application) tail of the distribution, \ie Value at Risk, and (ii) testing for non-zero lagged cross-correlations between the two binary time series representing the occurrences of extreme left (right) tail events, with a method based on spectral analysis. Based on this test, \eg, \cite{pirino} have studied the network of causal relations detected by Granger causality in tail for a bipartite financial system of banks and sovereign bonds and, combining measures of network connectedness with the ratings of the sovereign bonds, proposed a flight-to-quality indicator to identify periods of turbulence in the market. However, as we show below, the statistical test of Granger causality in tail by \cite{hongetal} displays some sensitivity to both non-zero auto-correlation and instantaneous cross-correlation in the binary time series representing extreme events, resulting in an increased rate of false causality detections. Additionally, the test by \cite{hongetal} is by construction a pairwise causality analysis, thus sensitive to false detections when variables of some importance, different from the two under investigation, are not considered. 

In this paper we propose a different approach to identify Granger causality in tail, which overcomes some of the issues of the Hong et al. method. Differently from the latter, our approach is parametric and %The statistical test for Granger causality in tail we propose is based on an explicit model of the causation between extreme events. The advantage of the parametric approach is that we
it explicitly models causality interactions between time series. Thus, it is less sensitive to other possible effects, such as autocorrelation, which may result in spurious detections as mentioned above, and can be generalized to multivariate settings. %This is in effect as in the original Granger test \cite{granger}. 
Moreover, having an explicit model of the causation process allows us to devise a method based on Likelihood-Ratio to test for Granger causality in tail. Specifically, the causation mechanism is captured by a process in which the extreme events are 
%copied from past extreme events 
modeled according to a {\it discrete autoregressive} process of order $p$, namely DAR(p).  The DAR(p) process, first introduced by \cite{dar}, is the natural extension of the standard autoregressive process for binary time series. We consider a multivariate generalization of the DAR(p) process, {namely VDAR(p)}, where the binary variable $X$ describing the occurrence of an extreme event for the underlying time series $x$ can be copied from its past values or from the past values of the extreme events of another time series $y$.
{The choice of this specific model is motivated by the fact that the Markov chain associated with the vector discrete autoregressive process of order one (VDAR(1)) can be interpreted as the maximum entropy distribution of binary random variables with given means,  lag one auto-correlations, and lag one cross-correlations, as recently proved in \cite{campajolaPRErapid}. Moreover, the same argument holds for the vector discrete autoregressive process of generic order, by noticing that a Markov chain of order $p$ for $N$ variables can be seen equivalently as a Markov chain of order one for $N\times p$ variables, under appropriate conditions for the transition matrix. By the principle of maximum entropy then the VDAR(p) model is the first candidate in building a parametric method to test for non-zero lagged cross-correlations (of binary time series) as signal of Granger causality in tail.}

We first propose a statistical test based on the bivariate version of the model, for any $p$. Then, to overcome the limit of pairwise analysis highlighted above, we also propose a statistical method for the multivariate case, with $p=1$ (Markovian dynamics). 
%Here, we propose a novel statistical test which accounts naturally for autocorrelated binary random variables and flexible enough to multivariate generalization. We introduce a statistical approach to test for Granger causality in tail by explicitly modeling causation between $x$ and $y$ with the mechanism of copying the occurrence of an extreme event for $x$ from the past realizations of extreme events of $y$. The mechanism of copying from the past is captured by the stochastic DAR(p) process, firstly introduced by \cite{dar}, namely the natural extension of the standard autoregressive process for a binary random variable. Here, we propose for the first time the multivariate generalization of the DAR(p) process, then introducing a method based on the Likelihood-Ratio for testing Granger causality in tail in the original spirit of \cite{granger}.

Our findings show that the detection of causality between extreme events is far from a trivial task, with a significant dependence on the adopted statistical procedure. In fact, we show that the proposed method and the current standard in literature represented by the test of \cite{hongetal} differ under some circumstances. First, as mentioned above, the test by Hong et al. displays some sensitivity to auto-correlation and cross-correlation of the time series of extreme events, which are, on the contrary, naturally accounted for by our framework. We show numerically concrete examples of such behavior and the spurious effect of two-way causality detection in the presence of unidirectional relations, a drawback of the Hong et al.'s test which is solved by our method. As a consequence, we claim that our method should be preferable in such cases. At the same time, we present also cases when the approach by Hong et al. outperforms ours, for instance when the underlying dynamics of the time series is autoregressive and heteroskedastic, even if the discrepancy is typically quite small. {In fact, in the case of model misspecification, a non-parametric approach should be preferable to a parametric method. Nevertheless, we show numerically also cases when the latter works better, even for misspecified data generating process.} Second, we point out the importance of a multivariate approach, numerically showing the consequences of network effects in the spurious detection of Granger causality in tail relations when adopting a pairwise analysis. We then propose a possible solution within our framework. Finally, in an empirical application to high-frequency price returns of a portfolio of stocks traded in the US stock markets, we highlight that different methods yield different networks of causal interactions between stocks. First, while  Hong et al.'s approach results in an almost complete graph, both the pairwise and multivariate versions of the statistical method we propose give much sparser networks. Hence, the measure of the level of causality in the system dynamics depends on the chosen approach. Additionally, the captured dynamics of the network evolution itself is quite different, in particular in relation to the presence of the financial crisis of 2007-2008: no patterns are recognized by using the test of Hong et al., while, with our method, a sharp transition characterizes the number of causal interactions between the stocks composing the financial sector and the others. In particular, we find that the financial sector started to be less `Granger-caused' by the other stocks before the financial crisis, but it `Granger-causes' more than the average over all the considered period. Thus, our findings open a discussion about the correct evaluation of a Granger causality relation between tail events and, in this paper, we highlight both advantages and drawbacks of the different approaches together with some signals associated with false detections.

The paper is organized as follows. Section \ref{sec:GCT} presents the general definition of Granger causality in tail and explains how to obtain the sequence of extreme events from data. We also discuss the importance of the multivariate approach to avoid false detections because of network effects. Section \ref{sec:method} introduces the novel methodology and describes how to construct the test statistics. Section \ref{sec:montecarlo} presents some Monte Carlo exercises to validate numerically the novel approach and to compare it with the test by \cite{hongetal}. Section \ref{sec:application} shows a financial application of the method. Finally, Section \ref{sec:conclusion} concludes. Technical details are reported in Appendix \ref{app:mle}.

\section{Granger causality in tail}\label{sec:GCT}
As introduced for the first time by \cite{granger}, the concept of {\it Granger causality} relies on testing whether the past information on a time series $y$ is statistically useful in predicting the future of another time series $x$, better than using only the past information on $x$. In the original version of the test, the information on past realization of the two time series defines the information set, which is also called the {\it Universe}. The information set may include also the information on other variables.

Here, we study Granger Causality (GC) {\it in tail}, that is a generalization of the standard Granger causality, but focusing on the prediction of {\it extreme} events represented by binary time series. With a similar aim of \cite{granger}, we define:
\begin{definition}{{\it (Granger Causality in tail)}.}\label{def1}
A time series $y$ `{\it Granger-causes in tail}' another time series $x$ if, given some information set, we are able to predict the occurrence of an {\it extreme} event for $x$ using the past information on {\it extreme} events of $y$ better than if the information on {\it extreme} events of $y$ is not considered.
\end{definition}

To characterize whether the occurrence of an extreme event for $y$ can help predicting the occurrence of an extreme event for $x$ in the spirit of Granger causality, we need to specify how an extreme event is defined, giving an operational meaning to definition \ref{def1}.
%from an operative point of view.

\subsection{Extreme events and Value at Risk}
Assume to observe a realization at time $t$ of a random variable $x$, \ie $x_t\in\mathbb{R}$. Given the distribution of the random variable conditional to some suitably defined information set $I_{t-1}$ up to time $t-1$\footnote{Here, we consider the stochastic process in discrete time with the unit value representing the time scale of observations.}, the realization $x_t$ is defined as {\it extreme} if it is in the left (or, equivalently right) tail of the distribution.

A natural way to characterize the tail of a distribution is the Value at Risk (VaR), see, \eg, \cite{roy}. In the case of the left tail, for a given probability level $\rho\in(0,1)$, the Value at Risk $VaR_{\rho,t}$ at time $t$ is the quantile of the distribution of $x$ conditional to the information set $I_{t-1}$ and associated with the probability $\rho$. It is implicitly defined by
$$
\P(x_t<VaR_{\rho,t}|I_{t-1})=\rho\:\:\:\mbox{almost surely (a.s.)}.
$$
Given the time series $\{x_t\}_{t=1,...,T}$  which describes the stochastic process for $x$ at discrete time, there exist several methods to estimate the Value at Risk of the distribution of $x$ at time $t$: Monte Carlo simulation methods, Hansen's autoregressive conditional density estimation \cite{hansen}, Morgan's RiskMetrics \cite{morgan}, historical estimation of the realized variance \cite{barndoff1},  and Engle and Manganelli's conditional autoregressive VaR (CAViaR) model \cite{englemanganelli}. 

In the case of financial time series, we can be interested, \eg, in extreme variations of prices, thus we can compare the price return $x_t$ with an estimate of the instantaneous or spot volatility $\sigma_t$, \ie
\be\nonumber
\frac{x_t}{\sigma_t}<\theta,
\ee
where the value of $\theta$ is either chosen as a free parameter or determined by the desired probability level $\rho$ of the Value at Risk, given some assumption on the probability distribution of returns. %The previous condition defines the occurrence of a (negative) extreme event for the price. 
The crucial aspects in this approach are the proper estimation of the spot volatility, see \cite{barndoff2,mancini,corsiTBV} and the choice of the conditional density (when one is interested in mapping $\theta$ to $\rho$).

Sometimes, the tail probabilities depends not only on the mean and the variance of the distribution, but also on other moments such as skewness and kurtosis, \eg see the empirical financial studies by \cite{harvey1,harvey2,jondeau}. Some econometric models can capture time-varying higher order conditional moments, such as \cite{gallant,hansen}. %However, more generally, note that VaR is well-defined even if the stochastic process is not covariance-stationary, \ie the second moment of the distribution of returns does not exist. In this case, a possible solution is given by the conditional autoregressive VaR (CAViaR) model of \cite{englemanganelli}, which captures directly the evolution of the Value at Risk.

Finally, given some estimation of the Value at Risk, we can map the time series of realizations $\{x_t\}_{t=1,...,T}$ of $x$ to the binary time series $\{X_t\}_{t=1,...,T}$ of extreme events of $x$, by defining $X_t=1$ if $x_t<VaR_{\rho,t}$\footnote{In the case of the right tail, it is $x_t>VaR_{1-\rho,t}$.}, $0$ otherwise.

\subsection{Null vs Alternative Hypotheses for GC in tail}
The statistical test of Granger causality in tail as defined in Def. \ref{def1} can be formalized as follows.

Given the times series of extreme events (or {\it hits}) associated with $x$ and $y$, \ie $X$ and $Y$, and all other information about the universe included in some information set $U$, the {\it null hypothesis} that $y$ does not `Granger-cause in tail' $x$ can be stated as
\be\label{nullH0}
\mathbb{H}^0: \P(X_t = 1|I_{t-1}^X,I_{t-1}^Y,U_{t-1}) = \P(X_t=1|I_{t-1}^X,U_{t-1})\:\:\:\mbox{a.s.}\:\:\:\forall t,
\ee
where $I_{t-1}^X\equiv\{X_s\}_{s=1,...,t-1}$, $I_{t-1}^Y\equiv\{Y_s\}_{s=1,...,t-1}$, and $U_{t-1}$ the set of all available information (on other time series, possibly) up to time $t-1$.

On the other hand, the alternative hypothesis is
\be\label{altHA}
\mathbb{H}^A:  \P(X_t = 1|I_{t-1}^X,I_{t-1}^Y,U_{t-1}) \neq \P(X_t=1|I_{t-1}^X,U_{t-1})\:\:\:\mbox{a.s.}\:\:\:\forall t.
\ee
We say that $y$ `Granger-causes in tail' $x$ if the null hypothesis $\mathbb{H}^0$ is rejected. Thus, under the alternative hypothesis $\mathbb{H}^A$, the information on the past extreme events of $y$, \ie $I_{t-1}^Y\equiv\{Y_s\}_{s=1,...,t-1}$, can be used to obtain a better prediction on the occurence of  an extreme event of $x$, \ie  $X_t$, with respect to the prediction obtained not accounting for it. %In this case, we say that $y$ `Granger-causes in tail' $x$.

%The operative definitions of the null hypothesis (\ref{nullH0}) together with the alternative hypothesis (\ref{altHA}) associated with the Definition \ref{def1} of Granger causality in tail are thus used to built the test statistics.

\subsection{Hong et al.'s test}\label{subsec:hongetal}
The method proposed by \cite{hongetal} is a kernel-based non-parametric test for Granger causality in tail, which is built on the null hypothesis (\ref{nullH0}) and having test statistics based on the normalized cross-spectral density between two time series $X$ and $Y$,
\be\label{hongteststat}
f(\omega)=\frac{1}{2\pi} \sum_{\tau=-\infty}^{+\infty}\rho(\tau)e^{-i\omega\tau}
\ee
with $\rho(\tau)\equiv\mbox{Corr}(X_t,Y_{t-\tau})$. In particular, under the null hypothesis (\ref{nullH0}), it is $\rho(\tau)=0$ $\forall \tau>0$. Thus, using spectral methods similarly to \cite{hong2001}, a test statistic can be defined to control for non-zero lagged cross-correlations between the two binary time series $\{X_t\}_{t=1,...,T}$ and $\{Y_t\}_{t=1,...,T}$. The statistical rejection of zero cross-correlation coefficients is thus a signal of a causality relation. This approach considers the possibility of a causal interaction at any possible time lag. However, because of finite sample size, the largest time scale is imposed in practice by the kernel estimator of the cross-spectral density (\ref{hongteststat}), in particular by the kernel non-uniform weighting whose effective window is fixed by an integer parameter $M$. In the following, we adopt the Daniell kernel which has displayed the best performance in the numerical analysis presented in \cite{hongetal}, with $M=5$ when not specified differently. 

\subsection{Network effects and multivariate analysis}
\label{sec:netw_eff}
As pointed out in the original paper of Granger causality \cite{granger}, the definition of the Universe is crucial. In the analysis of a causality relation between $x$ and $y$, the time series $\{x_t,y_t\}_{t=1,...,T}$ are considered as {\it all} available information. This assumption is  strong and unrealistic. In fact, consider the simplest case in which all relevant information is numerical in nature and refers to three stochastic processes $x$, $y$, and $z$. Now suppose to test the presence of a causal relation between $x$ and $z$, whereas the true causality chain is from $x$ to $y$, and from $y$ to $z$. Then, neglecting the information on $y$ when testing causality between $x$ and $z$ can induce a spurious causality detection. This is similar to spurious correlation between sets of data that arise when some other statistical variable of importance has not been included.

False detection of causal relations becomes relevant when we analyze a system displaying by nature a complex network of interactions between its many subparts. In this scenario, pairwise causality analysis applied to all possible pairs of elements will capture the network of causal relations, but also spurious effects. This could be the case, for example, of financial networks of investors \cite{billio} or of banks and bond investments \cite{pirino} where the degree of connectedness in the Granger causality network (for both in mean and in tail cases) is used to construct indicators of financial distress. Thus, reducing false detections is of paramount importance for the correct evaluation of such indicators. To this end, the causality analysis needs to be generalized to the multivariate case, thus accounting for the information on all subparts composing the whole system under investigation (but assuming that no important variables have been excluded).

Granger causality in mean has been extended to the multivariate case using the vector autoregressive process VAR(p) for more than two variables \cite{geweke} and several procedures have been introduced to validate the statistical significance of the off-diagonal couplings, \ie non-zero parameters capturing the presence of causal interactions, such as the {\it LASSO} penalization \cite{lasso} or methods based on the transfer entropy \cite{barrett}. In the next Section, after introducing a pairwise test of Granger causality {\it in tail} based on {\it Discrete AutoRegressive} DAR(p) processes \cite{dar} for binary time series, which we generalize to the bivariate case, we also present a Markovian ($p=1$) extension to the multivariate case with more than two variables, in the same spirit of both the milestone works of \cite{granger,geweke}.
%, by proposing the multivariate generalization of the {\it discrete autoregressive process} DAR(p) \cite{dar} for binary random variables, similarly to what is done for testing Granger causality {\it in mean} by using the {\it vector autoregressive} VAR(p) model for (real) random variables.

%\comment[id=PM]{prova commento}
\section{Methods and test statistics}\label{sec:method}
In this section we propose the multivariate generalization of the DAR(p) process to describe jointly the dependence structure of $N$ binary random variables which represent the extreme events of some underlying time series. The introduction of this modeling framework permits to test for the presence of non-zero terms of (lagged) interactions between the binary variables, which signal causal relations.

The discrete autoregressive model DAR(p) has been introduced for the first time by \cite{dar} and recently applied to the modeling of temporal networks \cite{williams1, mazzarisi, williams2}. In the most general setting, it describes the time evolution  of a categorical variable which has $p$-th order Markov dependence and a multinomial marginal distribution. Here we are interested in the case of a binary random variable $X$, thus the marginal distribution is Bernoulli. If $X_t$ is the realization of $X$ at time $t$, we have
\be\label{darp}
X_t=V_t\:X_{t-\tau_t}+(1-V_t)Z_t
\ee
where $X_t\in\{0,1\}\:\forall t$, $V_t\sim\mathcal{B}(\tilde{\nu})$ is a Bernoulli random variable with $\tilde{\nu}\in[0,1]$, $\tau_t\sim \mathcal{M}(\tilde{\gamma}_1,...,\tilde{\gamma}_p)$ is a multinomial random variable with $\tilde{\bm{\gamma}}\equiv\{\tilde{\gamma}_j\}_{j=1,...,p}$ such that $\sum_{i=1}^p\tilde{\gamma}_i=1$ and $Z_t \sim \mathcal{B}(\tilde{\chi})$ is a Bernoulli random variable with $\tilde{\chi}\in[0,1]$. In other words, at each time, $V_t$ determines whether $X_t$ is copied from the past or sampled from the marginal. When $X_t$ is copied from the past, the multinomial random variable $\tau_t$ selects the time lag and, accordingly, which past realization of $X$ we copy.

The process defined by (\ref{darp}) has the property that the autocorrelation at any lag is larger than or equal to zero. This is by construction, since $\tilde{\nu},\tilde{\gamma}_1,...,\tilde{\gamma}_p$ are non-negative definite. Moreover, the Yule-Walker equations associated with (\ref{darp}) are formally equivalent to the ones of the standard AR(p) process, see \cite{dar}.

We consider the generalization of the process (\ref{darp}) to the case of a multivariate (binary) time series $\X\equiv\{X_t^i\}_{t=0,1,...,T}^{i=1,...,N}$. With respect to the univariate DAR(p), the difference is that when $X_i^t$ is copied from the past it can be copied either from its own past values or from the past values of one of the other $N-1$ variables. Specifically, the binary random variable $X_i^t$ is copied from the past with probability $\nu_i$, and in this case it selects which variable $j$ to copy according to the probabilities $\lambda_{ij}$ and the time lag according to the multinomial  $\tau_{t}^{ij}$. Otherwise, it is sampled from its marginal Bernoulli distribution with parameter $\chi_i$. We can write the evolution of $X_i$ as  
\be\label{vdarp}
X_t^i=V_t^i\:X_{t-\tau_t^{iJ_t^i}}^{J_t^i}+(1-V_t^i)Z_t^i
\ee
where $X_t^i\in\{0,1\}$ $\forall i,t$, $V_t^i\sim\mathcal{B}(\nu_i)$ with $\nu_i\in[0,1]$ $\forall i$, $\tau_t^{ij}\sim \mathcal{M}(\gamma_{ij,1},...,\gamma_{ij,p})$ $\forall i,j$ with $\bm{\gamma}_{ij}\equiv\{\gamma_{ij,s}\}_{s=1,...,p}$ such that $\sum_{s=1}^p\gamma_{ij,s}=1$, $J_t^i\sim\mathcal{M}(\lambda_{i1},...,\lambda_{iN})$ is a multinomial random variable with $\bm{\lambda}_i\equiv\{\lambda_{ij}\}_{j=1,...,N}$ such that $\sum_{j=1}^N\lambda_{ij}=1$ $\forall i$, and $Z_t^i\sim\mathcal{B}(\chi_i)$ with $\chi_i\in[0,1]$, $\forall i$. For notational simplicity, let us define $\bm{\gamma}\equiv\{\bm{\gamma}_{ij}\}_{i,j=1,...,N}$ and $\bm{\lambda}\equiv\{\bm{\lambda_i}\}_{i=1,...,N}$. We refer to (\ref{vdarp}) as the {\it Vector Discrete AutoRegressive} VDAR(p) process.

%The process (\ref{vdarp}) describes the evolution of generic binary random variable $X_t^i$ with the mechanism of copying from the past with probability $\nu_i$, then choosing either $X^i$ itself or one of the other $N-1$ random variables, with probabilities $\lambda_{ij}$ $\forall j=1,...,N$, at past time lag selected according to $\tau_t^i$, otherwise sampling $X_t^i$ with Bernoulli marginal probability.

If the off-diagonal interaction term $\lambda_{ij}$ with $i\neq j$ is non-zero, a causal relation in the sense of Granger is present between $X^i$ and $X^j$, that is,  the occurrence of $X^j_t=1$ (an extreme event for the underlying time series $x^j$) increases the probability of observing one hit $X^i_{t+t'}=1$ at some future time $t+t'$ (depending on $\tau^{ij}$). Therefore, to test for Granger causality in tail between $X^i$ and $X^j$ we must assess the statistical significance of the off-diagonal term $\lambda_{ij}$ of the process (\ref{vdarp}). 

In principle, this approach allows to analyze the time scale at which a causal relation occurs by properly selecting the time lag associated with the mechanism of copying from the past. However, the number of parameters of the model (\ref{vdarp}) grows as $O(Np)$, thus some restrictions of the parameter space are needed for computational reasons. For this reason, in sections \ref{sec:pairwise} and \ref{sec:multivariate} we present two possible such restrictions to construct a statistical test: the bivariate case $N=2$, with $p>1$ and the multivariate Markovian case $p=1$ with $N>2$.
%Below, we consider the bivariate version of the process (\ref{vdarp}) in order to construct our statistical test of Granger causality in tail when only two binary random variables describing the occurrence of extreme events are considered. Then, we consider the Markovian process (\ref{vdarp}) with $p=1$ but generic $N>2$ to describe the case of the Universe including all variables at once.

\subsection{Pairwise causality analysis}
\label{sec:pairwise}
Pairwise analysis aims at detecting Granger causality in tail between two time series, and it excludes the information on any other variables. This approach is sufficient if, at least approximately,  no other variable affects the dynamics of the two considered ones. In this situation, in fact, we can ignore the risk of spurious detections discussed in section \ref{sec:netw_eff}. Let $\{X_t\}_{t=1,...,T}$ and $\{Y_t\}_{t=1,...,T}$ be the binary time series representing the occurrences of extreme events. We describe their dependence structure with the {\it bivariate} VDAR(p) process (\ref{vdarp}).

\subsubsection{Bivariate VDAR(p)} 
The bivariate version of the model (\ref{vdarp}) can be specified more explicitly as
\be\label{bidarp}
\begin{cases}
X_t &= V_t^1((1-J^1_t)X_{t-\tau_t^{11}}+J_t^1Y_{t-\tau_t^{12}})+(1-V^1_t)Z_t^1\\
Y_t &= V^2_t(J^2_tX_{t-\tau_t^{21}}+(1-J^2_t)Y_{t-\tau_t^{22}})+(1-V^2_t)Z_t^2
\end{cases}
\ee
where $X_t,Y_t\in\{0,1\}\ \forall t$, $V_t^i\sim\mathcal{B}(\nu_i)$ with $\nu_i\in[0,1]$ $\forall i=1,2$, $J^i_t\sim\mathcal{B}(\lambda_i)$ with $\lambda_i\in[0,1]$ $\forall i=1,2$, and $\tau_t^{ij}\sim\mathcal{M}(\gamma_{ij,1},...,\gamma_{ij,p})$ with $\sum_{s=1}^p\gamma_{ij,s}=1$. The marginals $Z_t^1$ and $Z_t^2$ are also Bernoulli random variables with distribution $\mathcal{B}(\chi_1)$ and $\mathcal{B}(\chi_2)$, respectively, with $\chi_{1},\chi_2\in[0,1]$.

The process (\ref{bidarp}) describes the evolution of the binary time series $X$, as follows. At time $t$, (i) $V_t^1$ determines whether $X_t$ is copied from the past (with probability $\nu_i$) or not; (ii) if it is, $J^1_t$ determines whether it is copied from a past value of $X$, $X_{t-\tau_{t}^{11}}$, (with probability $1-\lambda_i$) or  from a past value of $Y$,  $Y_{t-\tau_{t}^{12}}$ (with probability $\lambda_i$); (iii) the time lag is determined by the multinomial random variable $\tau_t^{11}$ (for $X_{t-\tau_{t}^{11}}$) or $\tau_t^{12}$ (for $Y_{t-\tau_{t}^{12}}$); (iv) if $X_t$ is not copied from the past, its value is 1 with probability $\chi_1$ and 0 otherwise. Equivalently for the binary time series $Y$. Hence, the parameter $\lambda_1$ controls the level of dependence of $X$ from $Y$ (and vice versa when considering $\lambda_2$): conditional on the probability that a past event affects the current realization (\ie $\nu_1>0$), the larger is $\lambda_1$, the larger is the probability that the occurrence of an extreme event $Y_{t-\tau_t^{12}}=1$ in the past triggers an extreme event $X_t=1$. If this is the case, taking into account the past information on $Y$ helps in forecasting $X$, thus there exists a causal relation from $Y$ to $X$.

We can test for Granger causality in tail by constructing a test statistic based on the Likelihood-Ratio as follows.

%\subsubsection{Bivariate-DAR(1)} Bi-DAR(1) model describes a bivariate time series of binary random variables $X$ and $Y$ whose values at time $t$ are
%\be\label{bidar1}
%\begin{cases}
%x_t &= v_t(a_tx_{t-1}+(1-a_t)y_{t-1})+(1-v_t)\varepsilon_t\\
%y_t &= s_t((1-b_t)x_{t-1}+b_ty_{t-1})+(1-s_t)\eta_t
%\end{cases}
%\ee
%where $x_t,y_t\in\{0,1\}\:\forall t$, $v_t\sim\mathcal{B(\nu)}$ with $\nu\in[0,1]$, $s_t\sim\mathcal{B(\xi)}$ with $\xi\in[0,1]$, $a_t\sim\mathcal{B(\alpha)}$ with $\alpha\in[0,1]$ and $b_t\sim\mathcal{B(\beta)}$ with $\beta\in[0,1]$. The marginal $\varepsilon_t$ ($\eta_t$) is also a Bernoulli random variable with distribution $\mathcal{B}(\chi)$ ($\mathcal{B}(\mu)$) with $\chi\in[0,1]$ ($\mu\in[0,1]$).

\subsubsection{Likelihood-Ratio (LR) statistic} In order to construct the statistical test for Granger causality in tail between two time series, we need to assess the statistical significance of the off-diagonal autoregressive coefficients of the bivariate VDAR(p) process (\ref{bidarp}). We propose to adopt the Likelihood-Ratio (LR) test \cite{hansenLR} by stating the null hypothesis (\ref{nullH0}) together with the alternative hypothesis (\ref{altHA}) in terms of the likelihood of two competing models, namely VDAR(p) and DAR(p).
%by exploiting similar considerations to 
This is further in line with \cite{barnett1,barnett2}, where authors notice that the GC (in mean) statistic may be formalized as a likelihood-ratio test.

The null hypothesis $\mathbb{H}^0$ (\ref{nullH0}) that the time series $Y$ {\it does not} Granger-cause in tail the time series $X$ can be stated as
\be\label{H0bidarp}
\begin{split}
\mathbb{H}^0:\: &\P^{DAR(p)}(X_t=1|\{X_{t-s}\}_{s=1,..,p},\tilde{\nu},\tilde{\bm{\gamma}},\tilde{\chi})\\&=\P^{VDAR(p)}(X_t=1|\{X_{t-s}\}_{s=1,..,p},\{Y_{t-s}\}_{s=1,..,p},\nu_1,\lambda_1,\bm{\gamma},\chi_1)\:\:\mbox{a.s.}
\end{split}
\ee
versus the alternative hypothesis $\mathbb{H}^A$ (\ref{altHA}) formulated as
\be\label{HAbidarp}
\begin{split}
\mathbb{H}^A:\: &\P^{DAR(p)}(X_t=1|\{X_{t-s}\}_{s=1,..,p},\tilde{\nu},\tilde{\bm{\gamma}},\tilde{\chi})\\&\neq\P^{VDAR(p)}(X_t=1|\{X_{t-s}\}_{s=1,..,p},\{Y_{t-s}\}_{s=1,..,p},\nu_1,\lambda_1,\bm{\gamma},\chi_1)\:\:\mbox{a.s.}
\end{split}
\ee
where $\P^{DAR(p)}(X_t=1|\{X_{t-s}\}_{s=1,..,p},\tilde{\nu},\tilde{\bm{\gamma}},\tilde{\chi})$ is the probability of an extreme event for the DAR(p) model (\ref{darp}), whereas $\P^{VDAR(p)}(X_t=1|\{X_{t-s}\}_{s=1,..,p},\{Y_{t-s}\}_{s=1,..,p},\nu_1,\lambda_1,\bm{\gamma},\chi_1)$ is for the VDAR(p) model (\ref{bidarp}).

Notice that the two considered models are nested, since the `full' VDAR(p) model (\ref{bidarp}) contains all the terms of the `restricted' DAR(p) model (\ref{darp}), but including also the `off-diagonal' terms of interaction. Thus, to make testable the null hypothesis (\ref{H0bidarp}), we can apply the likelihood-ratio test \cite{hansenLR} to assess the goodness of fit of the two competing nested models by evaluating how much better the full model works with respect to the restricted one: if the null $\mathbb{H}^0$ is supported by the observed data, the likelihoods of the two competing models should not differ by more than sampling error. Thus, we test whether the likelihood-ratio is significantly different from one, or equivalently whether its natural logarithm is significantly different from zero.

For notation simplicity, let us define $\theta_0\equiv\{\tilde{\nu},\tilde{\bm{\gamma}},\tilde{\chi}\}$ and $\theta\equiv\{\nu_1,\lambda_1,\bm{\gamma},\chi_1\}$, and indicate the likelihoods of the DAR(p) and VDAR(p) models as 
\be\label{likelihoodsmodels}
\begin{cases}
\mathcal{L}_0(\theta_0)&=\prod_{t=p+1}^T\P^{DAR(p)}(X_t|\{X_{t-s}\}_{s=1,..,p},\tilde{\nu},\tilde{\bm{\gamma}},\tilde{\chi}),\\
\mathcal{L}(\theta)&=\prod_{t=p+1}^T\P^{VDAR(p)}(X_t|\{X_{t-s}\}_{s=1,..,p},\{Y_{t-s}\}_{s=1,..,p},\nu_1,\lambda_1,\bm{\gamma},\chi_1).
\end{cases}
\ee
The likelihood-ratio is defined as 
\be\label{LR}
LR \equiv \frac{\sup_{\theta_0\in\Theta_0}\mathcal{L}_0(\theta_0)}{\sup_{\theta\in\Theta}\mathcal{L}(\theta)}= \frac{\mathcal{L}_0(\hat{\theta}_0)}{\mathcal{L}(\hat{\theta})}
\ee
where $\Theta_0$ and $\Theta$ represent the domains of the parameters $\theta_0$ and $\theta$, and $\hat{\theta}_0$ and $\hat{\theta}$ are the Maximum Likelihood Estimators (MLE) of the DAR(p) and VDAR(p) models, respectively. Therefore, in order to apply the test, we need to know the quantiles of the distribution of LR corresponding to the desired significance of the test and the MLE of the two models. \\
In most cases, the exact distribution of $LR$ is difficult to determine. However, assuming $\mathbb{H}^0$ is true, there is a fundamental result by Samuel S. Wilks \cite{wilks,casella}: as the sample size $T\rightarrow\infty$, the test statistics
\be\label{Lstatistics}
\Lambda\equiv-2(\log\mathcal{L}_0(\hat{\theta}_0)-\log\mathcal{L}(\hat{\theta})).
\ee
is asymptotically chi-squared distributed  with number of degrees of freedom equal to the difference in dimensionality of $\Theta$ and $\Theta_0$\footnote{In our case, this difference is equal to $p$, \ie the order of the discrete autoregressive process (\ref{bidarp}).}. Hence, this implies that we can calculate the test statistics $\Lambda$ given the data $\{X_t,Y_t\}_{t=0,1,...,T}$ for finite $T$, and then compare $\Lambda$ with the quantile of the chi-squared distribution corresponding to the desired statistical significance. We clearly expect that the larger is $T$, the better the test works.\footnote{{In fact, the convergence to the asymptotic $\chi^2$-distribution for the test statistic $\Lambda$ is mirrored by the convergence of the rejection rate under the null hypothesis to the adopted confidence level, usually considered equal to $5\%$. In our case, this is verified for very large sample size ($T\geq 5\times10^6$), thus pointing out the numerical consistency of the approach, despite the slow convergence rate. This behavior, however, opens to the possibility of considering other approaches, different from the Likelihood-Ratio test based on the Wilks' theorem, to define the test statistic of Granger causality in tail. For instance, a statistical test based on bootstrapping methods can be devised. This is left for future research.}}

The MLE of the  DAR(p) and VDAR(p) models is relatively simple to derive, see Appendix \ref{app:mle}. 
To improve the accuracy of the maximum likelihood estimation, it is convenient to use as a starting point of the algorithm the parameters obtained with the method of moments, namely the solution of the Yule-Walker equations for the DAR(p) and VDAR(p) models. In fact, Yule-Walker equations for DAR(p) and VDAR(p) are entirely equivalent to the ones of the autoregressive AR(p) and VAR(p) processes, respectively. The solution is a standard result of time series analysis, see \cite{tsay}. See Appendix \ref{app:mle} for the technical details about the inference process.

%\subsubsection{(Pairwise) Granger in tail network}
%
%description of how obtaining the causality network...

%\subsection{Out-of-sample analysis of Granger causality in tail}
%(a brief intro about the existence of out-of-sample Granger causality in mean and what is the idea of out-of-sample GC in tail, in particular we can relax the hypothesis of first-order Markovian behavior)
%$$
%\vdots
%$$
%With a similar aim of \cite{chao}, we propose to `test' for Granger causality in tail by studying the out-of-sample performance in forecasting the time series of extreme events for a binary variable. To this end, we need to specify two models: given the time series $X$ and $Y$ of extreme events, one model describes both the time series by accounting for a complete dependence structure between $X$ and $Y$, whereas the second model each time series independently from the other one. Hence, the forecasting performance of the two models are compared to assess the relevance of one series in forecasting the other one.
%
%
%
%\subsubsection{Testing for out-of-sample Granger causality in tail} The test of Diebold-Mariano \cite{dieboldmariano} cannot be applied here because the forecast is not a smooth function of model parameters. Then, we can adopt some binary classifier, \eg the ROC curve.
%

\subsection{Multivariate causality analysis} 
\label{sec:multivariate}
The aim of the multivariate causality analysis is to detect a Granger causality in tail relation between a pair of time series, but extending the information set to all the relevant variables of the system under investigation. This framework accounts for the situation in which a variable is affected by several other variables (network effect). Let the $N$ binary time series $\{X_t^i\}_{t=1,...,T}^{i=1,...,N}$ describe the sequences of extreme events. We describe their dependence structure at unit time lag with the Markovian VDAR(1) model (\ref{vdarp}) with $p=1$. The Markovian restriction, as mentioned before, is necessary for computational reasons.

The VDAR(1) model is specified as
\be\label{vdar1}
X_t^i=V_t^i\:X_{t-1}^{J_t^i}+(1-V_t^i)Z_t^i
\ee
where $X_t^i\in\{0,1\}\:\forall i,t$, $V_t^i\sim\mathcal{B}(\nu_i)$ with $\nu_i\in[0,1]$ $\forall i=1,...,N$, $J_t^i\sim\mathcal{M}(\lambda_{i1},...,\lambda_{iN})$ is a multinomial random variable with $\bm{\lambda}_i\equiv\{\lambda_{ij}\}_{j=1,...,N}$ such that $\sum_{j=1}^N\lambda_{ij}=1$ $\forall i=1,...,N$, and $Z_t^i\sim\mathcal{B}(\chi_i)$ with $\chi_i\in[0,1]$, $\forall i=1,...,N$.

The generic entry $\lambda_{ij}$ of the matrix $\bm{\lambda}\equiv\{\lambda_{ij}\}_{i,j=1,...,N}$ determines the level of interaction between $X^i$ and $X^j$ if $i\neq j$, or self-interaction if $i=j$ (diagonal elements). Hence, the presence of a Granger causality relation from $X^j$ to $X^i$ in the multivariate case can be tested by validating non-zero $\lambda_{ij}$.

In Statistical Inference, there exist many regularization methods that force the estimation algorithm to infer a less complex model by putting some parameters to zero, when not statistically significant. The two most widely used types of regularization are the so-called L1 (\ie {\it LASSO}) and L2 regularization \cite{tibi}. Recently, Decimation \cite{decimation} has been proposed to infer the topology of the interaction network in models with pairwise interactions between binary random variables. The Decimation method has proved to be very efficient in recovering the network of interactions for a specific logistic regression model, namely the Ising model, in both static \cite{decimation} and kinetic \cite{decimationKIM,campajola,campajolaQF} cases, by setting the weakest interaction couplings to zero iteratively. Here, we adopt the Decimation method to validate the entries of the matrix $\bm{\lambda}$ estimated by maximum likelihood (see Appendix \ref{app:mle} for the technical details). The validated off-diagonal couplings constitute the detected causality relations.

\section{Monte Carlo simulations}\label{sec:montecarlo}
In this section we use Monte Carlo simulations to analyze the performance of the proposed methods and to compare it with  \cite{hongetal}. A first analysis (section \ref{sec:bidar}) is performed simulating two time series using the bivariate VDAR(p) (\ref{bidarp}) as data generating process and then testing for a causality relation with the two methods. The two methods' performances are compared on the basis of their False Positive Rate (FPR) and True Positive Rate (TPR). Then, in section \ref{sec:Garch}, we perform the same analysis using alternative data generating processes, specifically the GARCH-type models (adopted in \cite{hongetal}) and the bivariate VDAR(1) with non-independent marginals, to verify the numerical consistency of the proposed {\it parametric} Likelihood-Ratio test of Granger causality in tail even when the data generating process is not correctly specified. In these pairwise settings, we point out the drawbacks of the method of \cite{hongetal} which are naturally solved by using the proposed approach. Then, in section \ref{sec:Vdar} we move to a setting in which the pairwise approximation is not correct, by simulating time series with a multivariate VDAR(1) process. We show that the multivariate generalization of our method does not give rise to the spurious effects that are instead found by the \cite{hongetal} method in this case.

\subsection{Bivariate VDAR(p) as data generating-process}
\label{sec:bidar}
In the bivariate VDAR(p) process (\ref{bidarp}), a causal interaction from $Y$ to $X$ is present when $\lambda_1>0$. Therefore, the null hypothesis $\mathbb{H}^0$ in (\ref{H0bidarp}) is true when $\lambda_1 = 0$, while the alternative hypothesis $\mathbb{H}^A$ in (\ref{HAbidarp}) is true when $\lambda_1>0$. We generated time series of $X$ and $Y$ of length $T$ for different values of $T$ with the bivariate VDAR(p) model for $p=1,2$ and different values of $\lambda_1$. Table \ref{Tablenullbidarp} reports the rejection rate of $\Lambda$ in (\ref{Lstatistics}) for the proposed test of Granger causality in tail, in particular the False Positive Rate (FPR), \ie the percentage of rejections of the null hypothesis when $\lambda_1=0$, and the True Positive Rate (TPR), \ie the percentage of rejections of the null hypothesis when $\lambda_1>0$. The Likelihood-Ratio (LR) test has appropriate size for all $T$, \ie  FPR is below the $5\%$ significance level. It also has a good power under the alternative hypothesis $\mathbb{H}^A$ in (\ref{HAbidarp}), \ie it is successful at recognizing true causality when $\lambda_1>0$, and it becomes more powerful as $T$ increases. The TPR, as expected, increases with the degree of causal interaction (\ie $\lambda_1$), that is, interactions are increasingly detected when they are stronger.

{\footnotesize
\begin{center}
\begin{table}
{\begin{flushleft} Size and power at the $5\%$ significance level of the proposed test of GC in tail.\end{flushleft}}
 \begin{tabular}{||c c c c c c c c c||} 
 \hline
 \thead{DGP:\\VDAR(1)} & \thead{FPR\\ $\lambda_1=0$} & \thead{TPR\\ $\lambda_1=0.01$} & \thead{TPR\\ $\lambda_1=0.025$} &\thead{TPR\\ $\lambda_1=0.05$} & \thead{TPR\\ $\lambda_1=0.1$} & \thead{TPR\\ $\lambda_1=0.25$} & \thead{TPR\\ $\lambda_1=0.5$} & \thead{TPR\\ $\lambda_1=0.75$} \\ [0.5ex] 
 \hline\hline
$T=500$ & 0.02 & 0.04 & 0.08  & 0.11 & 0.24 & 0.57 & 0.94 & 1.00 \\ 
 \hline
$T=1000$ & 0.02 & 0.03 & 0.08 & 0.17 & 0.36 & 0.89 & 1.00 & 1.00\\ 
 \hline
$T=2000$ & 0.02 & 0.04 & 0.10  & 0.25 & 0.60 & 0.99 & 1.00 & 1.00\\ 
 \hline
 $T=5000$ & 0.02 & 0.05 & 0.20  & 0.49 & 0.92 & 1.00 & 1.00 & 1.00\\ 
 \hline
$T=10000$ & 0.03 & 0.10 & 0.33  & 0.79 & 0.99 & 1.00 & 1.00 & 1.00\\ [1ex] 
 \hline
\end{tabular}

\vspace{0.2cm}

\begin{tabular}{||c c c c c c c c c||} 
 \hline
 \thead{DGP:\\VDAR(2)} & \thead{FPR\\ $\lambda_1=0$} & \thead{TPR\\ $\lambda_1=0.01$} & \thead{TPR\\ $\lambda_1=0.025$} &\thead{TPR\\ $\lambda_1=0.05$} & \thead{TPR\\ $\lambda_1=0.1$} & \thead{TPR\\ $\lambda_1=0.25$} & \thead{TPR\\ $\lambda_1=0.5$} & \thead{TPR\\ $\lambda_1=0.75$} \\ [0.5ex] 
 \hline\hline
$T=500$ & 0.01 & 0.02 & 0.03  & 0.05 & 0.09 & 0.32 & 0.71 & 0.92 \\ 
 \hline
$T=1000$ & 0.02 & 0.03 & 0.05 & 0.06 & 0.19 & 0.68 & 0.93 & 0.99\\ 
 \hline
$T=2000$ & 0.01 & 0.02 & 0.05  & 0.12 & 0.40 & 0.94 & 1.00 & 1.00\\ 
 \hline
 $T=5000$ & 0.01 & 0.02 & 0.10  & 0.26 & 0.76 & 1.00 & 1.00 & 1.00\\ 
 \hline
$T=10000$ & 0.01 & 0.05 & 0.14  & 0.51 & 0.96 & 1.00 & 1.00 & 1.00\\ [1ex] 
 \hline
\end{tabular}
\caption{False Positive Rate (FPR) and True Positive Rate (TPR) of the test statistic $\Lambda$ (\ref{Lstatistics}) under the null $\mathbb{H}^0$ that $Y$ `does not Granger cause' $X$ in (\ref{H0bidarp}) with data generated according to the VDAR(p) model (\ref{bidarp}) for different values of $\lambda_1$ and different sample sizes $T$. The parameters of the VDAR(p) model are: $\nu_1,\nu_2=0.5$, $\chi_1,\chi_2=0.05$, $\lambda_2=0$, and $\gamma_{ij}=0.5$ $\forall i,j=1,2$ in the case $p=2$. Notice that data are generated with $p=1$ (above) or $p=2$ (below), but in obtaining the test statistic $\Lambda$ (\ref{Lstatistics}) $p$ is inferred according to the Bayesian Information Criterion (see Appendix \ref{app:mle}), thus no prior information on the time lag order is used when applying the proposed test. 
Each rejection rate is computed over a sample of $500$ seeds. }
\label{Tablenullbidarp}
\end{table}
\end{center}}

{\footnotesize
\begin{center}
\begin{table}
{\begin{flushleft} Size and power at the $5\%$ significance level of the test of GC in tail by \cite{hongetal}.\end{flushleft}}
 \begin{tabular}{||c c c c c c c c c||} 
 \hline
 \thead{DGP:\\VDAR(1)} & \thead{FPR\\ $\lambda_1=0$} & \thead{TPR\\ $\lambda_1=0.01$} & \thead{TPR\\ $\lambda_1=0.025$} &\thead{TPR\\ $\lambda_1=0.05$} & \thead{TPR\\ $\lambda_1=0.1$} & \thead{TPR\\ $\lambda_1=0.25$} & \thead{TPR\\ $\lambda_1=0.5$} & \thead{TPR\\ $\lambda_1=0.75$} \\ [0.5ex]
 \hline\hline
$T=500$ & 0.13 & 0.18 & 0.23  & 0.30 & 0.47 & 0.75 & 0.96 & 1.00 \\ 
 \hline
$T=1000$ & 0.20 & 0.19 & 0.27 & 0.38 & 0.55 & 0.94 & 1.00 & 1.00\\ 
 \hline
$T=2000$ & 0.21 & 0.23 & 0.31  & 0.43 & 0.75 & 0.99 & 1.00 & 1.00\\ 
 \hline
 $T=5000$ & 0.19 & 0.28 & 0.41  & 0.67 & 0.95 & 1.00 & 1.00 & 1.00\\ 
 \hline
$T=10000$ & 0.23 & 0.32 & 0.51  & 0.87 & 0.99 & 1.00 & 1.00 & 1.00\\ [1ex] 
 \hline
\end{tabular}

\vspace{0.2cm}

\begin{tabular}{||c c c c c c c c c||} 
 \hline
 \thead{DGP:\\VDAR(2)} & \thead{FPR\\ $\lambda_1=0$} & \thead{TPR\\ $\lambda_1=0.01$} & \thead{TPR\\ $\lambda_1=0.025$} &\thead{TPR\\ $\lambda_1=0.05$} & \thead{TPR\\ $\lambda_1=0.1$} & \thead{TPR\\ $\lambda_1=0.25$} & \thead{TPR\\ $\lambda_1=0.5$} & \thead{TPR\\ $\lambda_1=0.75$} \\ [0.5ex] 
 \hline\hline
$T=500$ & 0.13 & 0.14 & 0.17  & 0.18 & 0.34 & 0.67 & 0.94 & 0.99 \\ 
 \hline
$T=1000$ & 0.19 & 0.18 & 0.23 & 0.29 & 0.41 & 0.88 & 0.99 & 1.00\\ 
 \hline
$T=2000$ & 0.20 & 0.22 & 0.27  & 0.33 & 0.63 & 0.98 & 1.00 & 1.00\\ 
 \hline
 $T=5000$ & 0.18 & 0.21 & 0.30  & 0.51 & 0.90 & 1.00 & 1.00 & 1.00\\ 
 \hline
$T=10000$ & 0.20 & 0.28 & 0.40  & 0.73 & 0.98 & 1.00 & 1.00 & 1.00\\ [1ex] 
 \hline
\end{tabular}
\caption{False Positive Rate (FPR) and True Positive Rate (TPR) associated with the test by \cite{hongetal} under the null $\mathbb{H}^0$ that $Y$ `does not Granger cause' $X$ with data generated according to the VDAR(p) model (\ref{bidarp}) for different values of $\lambda_1$ and different sample sizes $T$. The test statistic is computed by using the Daniell kernel with $M=5$. The parameters of the VDAR(p) model are: $\nu_1,\nu_2=0.5$, $\chi_1,\chi_2=0.05$, $\lambda_2=0$, and $\gamma_{ij}=0.5$ $\forall i,j=1,2$ in the case $p=2$. Each rejection rate is computed over a sample of $500$ seeds.}
\label{Tablenullbidarp_hong}
\end{table}
\end{center}}
\begin{figure}
\includegraphics[width=0.49\textwidth]{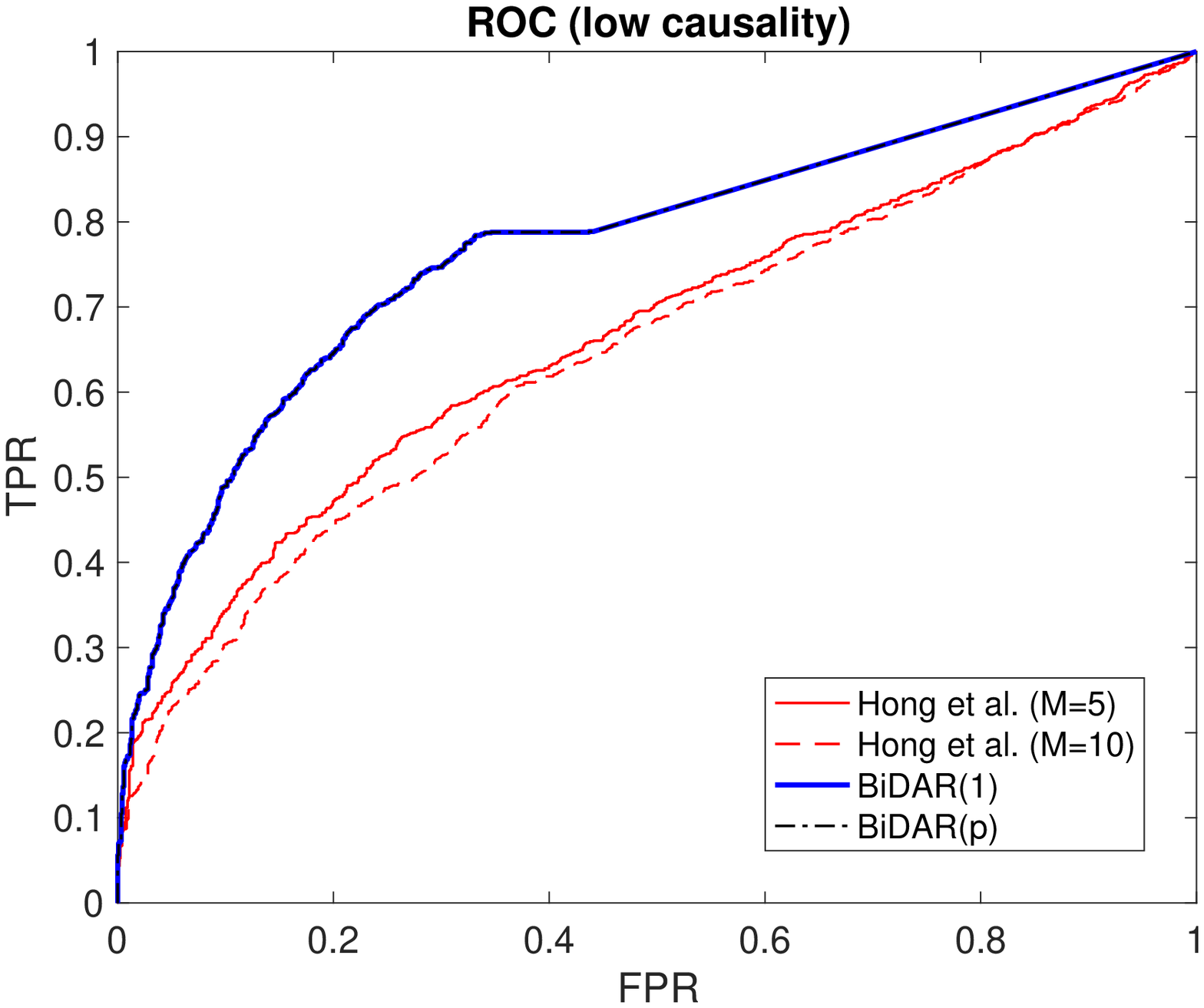}
\includegraphics[width=0.49\textwidth]{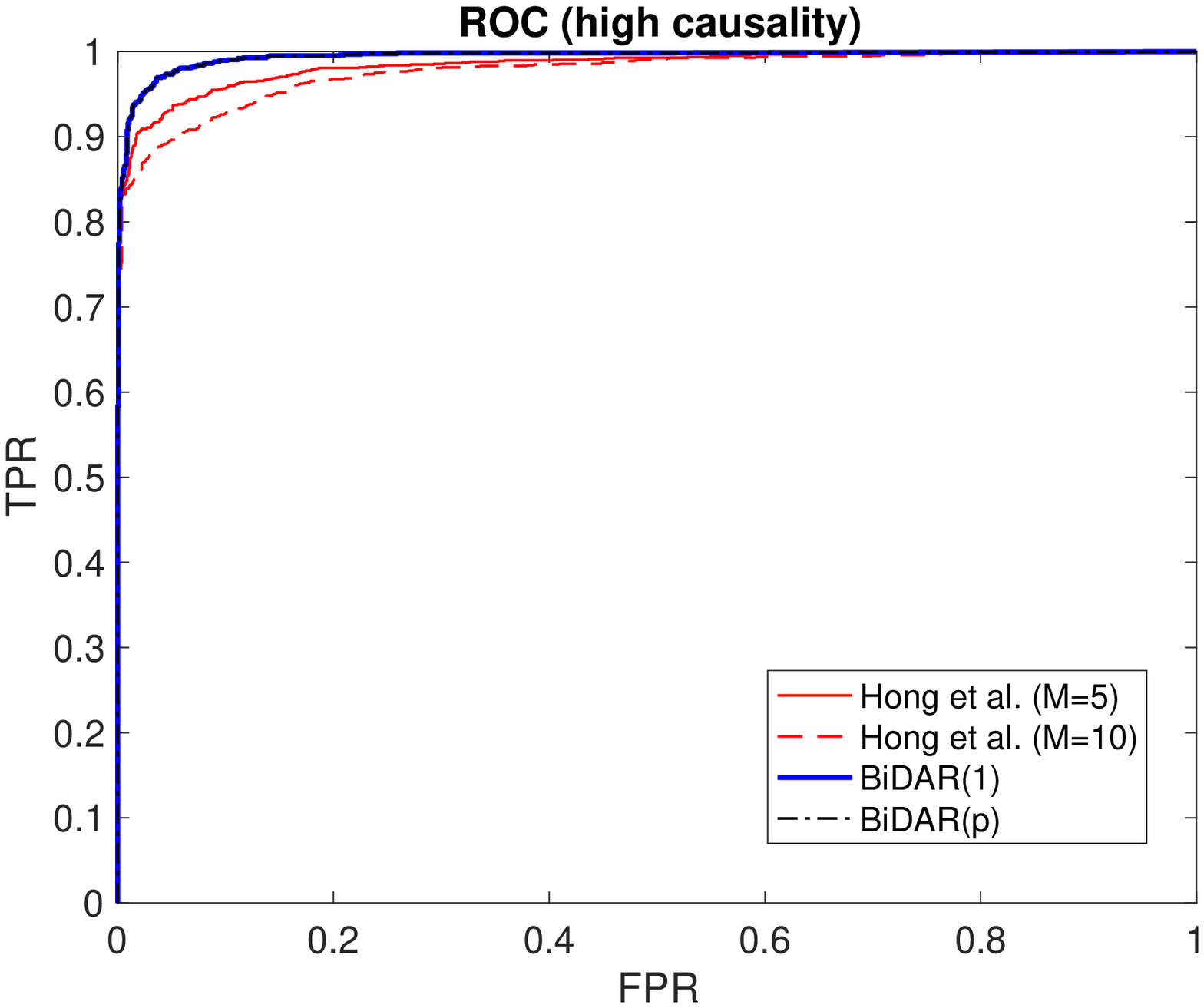}
\label{fig_comp1}
\caption{Receiving Operating Characteristic (ROC) curves built with the p-values associated with the test statistics of either the Likelihood-Ratio (LR) method (\ref{Lstatistics}) and the kernel-based non-parametric approach of \cite{hongetal}, used as binary classifiers depending on some threshold value, \ie the test significance level. Data are generated by the bivariate VDAR(1) model with $T=10000$, under the null $\mathbb{H}^0$ in (\ref{H0bidarp}) or under the alternative hypothesis $\mathbb{H}^A$ in (\ref{HAbidarp}), with one half probability over a sample of $5000$ simulations. The adopted model parameters are: $\nu_1=0.1$, $\nu_2=0$, $\chi_1,\chi_2=0.05$, $\lambda_2=0$, and $\lambda_1=0$ under the null $\mathbb{H}^0$, $\lambda_1\sim \mathcal{U}(0,0.25)$ (left) or $\lambda_1\sim \mathcal{U}(025,0.75)$ (right) under the alternative hypothesis $\mathbb{H}^A$. In implementing the LR test, the blue curve is obtained by imposing the order $p=1$ of the autoregressive process, whereas the black dotted curve (hidden under the blue curve) is obtained by optimally selecting the order $p$ with the Bayesian Information Criterion (see Appendix \ref{app:mle}). The method by Hong et al. is implemented with the Daniell kernel with a $M=5$ (red solid line) or $M=10$ (red dotted line).}
\label{fig_comp_bidarp_hong}
\end{figure}

We repeated the same analysis using the test of Granger causality in tail by \cite{hongetal}, see Table \ref{Tablenullbidarp_hong}. The test of Hong et al. tends to over-reject significantly the null hypothesis at the $5\%$ level, when data are generated with zero causal interaction, \ie FPR is much larger than $0.05$. This effect seems independent from the sample size $T$. At the same time, it displays a good power under the alternative hypothesis for $\lambda_1>0$, and the performances become better as $T$ increases. 

We remark that, for a given value of the significance level, the method by Hong  et al. shows a higher TPR with respect to our method (for fixed $\lambda_1$), but at the expense of a higher FPR. A fairer comparison between the two methods can be performed building the Receiving Operating Characteristic (ROC) curve associated with the two statistical methods, which allows to compare the TPRs of the two methods for a given FPR (Fig. \ref{fig_comp_bidarp_hong}). The curve corresponding to one method is built by plotting the points (FPR, TPR) obtained with the corresponding method for varying threshold values (\ie significance levels). We therefore simulated data according to the VDAR(1) process (details in the caption of Fig. \ref{fig_comp_bidarp_hong}), both with and without causality, and computed the FPR and TPR obtained by the two methods, resulting in the curves in Figure \ref{fig_comp_bidarp_hong}. For a given FPR, the best causality method displays the largest TPR. The results are shown in Figure \ref{fig_comp_bidarp_hong} for the two cases of low or high causality. In the low causality case, time series with causality are generated by the VDAR(1) model with $\lambda_1\sim \mathcal{U}(0,0.25)$, while in the high causality case with $\lambda_1\sim \mathcal{U}(0.25,0.75)$. In both cases, we can notice that the novel approach outperforms the non-parametric method by \cite{hongetal}. It is worth noting that we are able to identify correctly the time scale of the causal interaction, as evident from the superposition of the ROC curves built for VDAR(1) and VDAR(p) models: in the first case, the order of the autoregressive process is imposed by hand, whereas in the second case the order $p$ is optimally selected (\ie $p=1$) by the Bayesian Information Criterion during the Maximum Likelihood Estimation process, as explained in Appendix \ref{app:mle}.

{\footnotesize
\begin{center}
\begin{table}
{\begin{flushleft} Size at the $5\%$ significance level of the test of GC in tail by Hong et al. \cite{hongetal}.\end{flushleft}}
 \begin{tabular}{||c c c c c c c c||} 
 \hline
  \thead{DGP:\\VDAR(1)} & \thead{FPR\\ $\nu_2=0$} & \thead{FPR\\ $\nu_2=0.05$} & \thead{FPR\\ $\nu_2=0.25$} &\thead{FPR\\ $\nu_2=0.3$} & \thead{FPR\\ $\nu_2=0.4$} & \thead{FPR\\ $\nu_2=0.5$} & \thead{FPR\\ $\nu_2=0.75$}\\ [0.5ex] 
 \hline\hline
$M=5$ & 0.07 (0.02) & 0.07 (0.03) & 0.15 (0.02) & 0.36 (0.02) & 0.87 (0.02) & 0.99 (0.01) & 1.00 (0.01) \\ 
 \hline
$M=10$ & 0.06 (0.02) & 0.06 (0.02) & 0.16 (0.02) & 0.31 (0.02) & 0.84 (0.01) & 0.99 (0.02) & 1.00 (0.03)\\ 
 \hline
$M=15$ & 0.06 (0.02) & 0.07 (0.01) & 0.16 (0.02) & 0.27 (0.02) & 0.80 (0.02) & 0.99 (0.01) & 1.00 (0.01)\\ 
 \hline
 $M=20$ & 0.08 (0.02) & 0.08 (0.02) & 0.16 (0.02) & 0.30 (0.02) & 0.77 (0.02) & 0.99 (0.02) & 1.00 (0.01) \\ [1ex] 
 \hline
\end{tabular}
\caption{False Positive Rate (FPR) obtained for the test by \cite{hongetal} under the null $\mathbb{H}^0$ that $X$ `does not Granger cause' $Y$ with data generated according to the VDAR(1) model (\ref{bidarp}) for different values of $\nu_2$ and sample sizes $T=10000$. The test statistic is computed by using the Daniell kernel with different values of $M$. The parameters of the VDAR(1) model are: $\nu_1=\nu_2$, $\chi_1,\chi_2=0.05$, $\lambda_1=0.5$, and $\lambda_2=0$. Each rejection rate is computed over a sample of $500$ seeds. Values in the brackets represent instead the FPR of the LR test based on the test statistic (\ref{Lstatistics}).}
\label{Tablenullbidarp_hong_fp}
\end{table}
\end{center}}

Now, we show that when there is a unidirectional causal relation between two time series (e.g., $Y$ causes $X$ but not the converse), in the presence of non-zero autocorrelation the test by Hong et al. shows a very high FPR due to the mistaken detection of the inverse causal relation (from $X$ to $Y$), while our LR test has a small FPR rate. To show this,  we consider data generated by the VDAR(1) model (\ref{bidarp}) with $\lambda_1 = 0.5$ ($Y$ Granger-causes $X$) but $\lambda_2 = 0$ ($X$ {\it does not} Granger cause $Y$), and varying the parameter $\nu_2$ which determines how much the binary time series of $Y$ is autocorrelated. We then test for a Granger causality from $X$ to $Y$ with the method by Hong et al. Table \ref{Tablenullbidarp_hong_fp} shows the results of the numerical exercise. The rate of false rejections for the test of Hong et al. is increasing as the autocorrelation of binary time series increases, eventually converging to one.  For the novel LR test based on the VDAR(p) model, instead, the rate is below the level of the test significance, \ie $5\%$ (see the values in the brackets in Table \ref{Tablenullbidarp_hong_fp}). The poor performance of the non-parametric method by \cite{hongetal} can be understood by the following argument. The method tests for non-zero lagged cross-correlations between the two binary time series. Now, assume that the series are generated according to the VDAR(1) model with $\lambda_1>0$, $\lambda_2=0$, $\nu_2>0$. The fact that $\lambda_1>0$ implies, clearly, a non-zero covariance  $\E(\tilde{X}_t\tilde{Y}_{t-1})\neq 0$, where the tilde is for indicating the mean subtracted variable. However, since $Y$ is autocorrelated ($\nu_2>0$), this will also imply a non-zero covariance
\be
\begin{split}
\E(\tilde{Y}_t\tilde{X}_{t-1}) = &\: \E(V^2_t(J^2_t\tilde{X}_{t-1}\tilde{X}_{t-1}+(1-J^2_t)\tilde{Y}_{t-1}\tilde{X}_{t-1})+(1-J^2_t)\tilde{Z}_t^2\tilde{X}_{t-1}) = \\ 
= &\: \nu_2 \E(\tilde{X}_{t-1}\tilde{Y}_{t-1}) = \\
= &\: \nu_2 \E(V^2_{t-1}(J^2_{t-1}\tilde{X}_{t-1}\tilde{X}_{t-2}+(1-J^2_{t-1})\tilde{X}_{t-1}\tilde{Y}_{t-2})+(1-J^2_{t-1})\tilde{X}_{t-1}\tilde{Z}_{t-1}^2)=\\
= & \: (\nu_2)^2 \E(\tilde{X}_{t-1}\tilde{Y}_{t-2})\neq 0,
\end{split}
\ee
where we used $\E(\tilde{X}_t)=0$, $\E(\tilde{Z}_t^2)=0$, $\E(V^2_t)=\nu_2$, and $\E(J^2_t)=0$. Therefore, the method by Hong et al. tends to falsely detect causation from $X$ to $Y$ because of this non-zero lagged cross-correlation. This spurious effect is increasing with the autocorrelation itself. Our novel approach, on the contrary, is able to capture {both cross- and auto- correlations between binary time series, separately,} thus validating the correct direction of the causality relation (within the significance level).

{\footnotesize
\begin{center}
\begin{table}
 \begin{tabular}{||c c c ||} 
 \hline
 \thead{DGP:\\VDAR(1)} & $\mathbb{H}^0$ & $\mathbb{H}^A$  \\ [0.5ex] 
 \hline\hline
$\chi_{1,2} = 3\times10^{-4}$ & 0.001 &  0.831  \\
 \hline
$\chi_{1,2} =2.6\times 10^{-3} $ & 0.013 &  0.958  \\
 \hline
$\chi_{1,2} =8.6\times10^{-2} $ & 0.020 &  0.986  \\ [1ex] 
 \hline
\end{tabular}
\caption{Size and power at the $5\%$ significance level of the LR test of GC in tail, with statistic $\Lambda$ (\ref{Lstatistics}) under the null $\mathbb{H}^0$ that $Y$ `{\it does not} Granger cause' $X$, with data generated according to the VDAR(1) model for $T=98000$ with $\nu_1,\nu_2\sim U([0,1])$, $\lambda_1,\lambda_2=0$, and $\chi_1,\chi_2$ as indicated. Then, under the alternative hypothesis $\mathbb{H}^A$ that $Y$ `{\it does} Granger cause' $X$, data are generated with $\nu_1,\nu_2,\lambda_1,\lambda_2\sim U([0,1])$ and $\chi_1,\chi_2$ as indicated. Each rejection rate is computed over a sample of $1000$ seeds.}
\label{tabEmpiricalChi}
\end{table}
\end{center}}
%In summary, the numerical analysis to answer this point is reported in Table \ref{tabEmpiricalChi}, where we show the size and the power of the proposed test of Granger causality in tail for time series of length $T=98000$ (similarly to the empirical ones) generated by the VDAR(1) model, with parameters $\chi_1,\chi_2$ associated with the Bernoulli marginal distribution determining the frequency of tail events, taking the minimum, the mean, and the maximum values observed in empirical data. In all cases, we can notice that the Likelihood-Ratio test has appropriate size under the null $\mathbb{H}^0$ (below the $5\%$ significance level), and good power under the alternative hypothesis $\mathbb{H}^A$.

{To conclude this section, we repeat a similar Monte Carlo exercise for the Likelihood-Ratio test, but with data generated by the VDAR(1) model with sample size $T=98,000$ and three specific values for the $\chi_1,\chi_2$ parameters. This is done in view of the empirical application below, where financial time series of extreme events display often a very low frequency of tail events, thus associated with small $\chi_1$ and $\chi_2$. In particular, in Table \ref{tabEmpiricalChi} we report the size and power of the LR test associated with the minimum, the mean, and the maximum values of $\chi_1,\chi_2$ observed in empirical data. In all cases, we can notice that the LR test has appropriate size under the null $\mathbb{H}^0$ (below the $5\%$ significance level), and good power under the alternative hypothesis $\mathbb{H}^A$.}

\subsection{Alternative data generating process}
\label{sec:Garch}
We repeat the Monte Carlo exercise for pairwise causality analysis with the data generating process adopted in \cite{hongetal}. They consider a GARCH-type model for the underlying time series $x_1$ and $x_2$, from which the binary time series of extreme events are extracted. The model is the following:
\be\label{garch}
\begin{cases}
&x_{i,t}=\beta_{i1}x_{1,t-1}+\beta_{i2}x_{2,t-1}+u_{i,t},\:\:i=1,2 \\
& u_{i,t} = \sigma_{i,t}\epsilon_{i,t}\\
&\sigma_{i,t}^2=\gamma_{i,0}+\gamma_{i,1}\sigma_{i,t-1}^2+\gamma_{i,2}u_{1,t-1}^2+\gamma_{i,3}u_{2,t-1}^2\\
&\epsilon_{i,t}\sim N(0,1),\:\:\:i.i.d.\:r.v.\:\forall i=1,2,
\end{cases}
\ee
where the variance of $x_1$ and $x_2$ is described by a GARCH(1,1) process, but including also off-diagonal dependencies through the mixed terms $u_2^2$ and $u_1^2$, respectively for $x_1$ and $x_2$. Then, the realization at time $t$ of the process is the sum of the innovation term (as in the standard GARCH(1,1)) plus a linear combination of the past realizations at time $t-1$ of the two processes (similarly to standard vector autoregressive VAR(1) model). The following parameterization is used
\be\label{garch_par}
\begin{cases}
& (\beta_{11},\beta_{12},\gamma_{10},\gamma_{11},\gamma_{12},\gamma_{13})=(0.5,b,0.1,0.6,0.2,c)\\
& (\beta_{21},\beta_{22},\gamma_{20},\gamma_{21},\gamma_{22},\gamma_{23})=(0,0.5,0.1,0.6,0,0.2).
\end{cases}
\ee
In particular, the authors consider the following cases: NULL (no Granger causality in tail) when $b=c=0$; ALTER1 (Granger causality in tail from mean) $b=2, c=0$; ALTER2 (Granger causality in tail from variance) when $b=0, c=0.7$. 

For each individual time series, the authors propose to estimate by means of the Quasi-MLE method the following GARCH-type model (without off-diagonal interaction terms)
\be\label{garch0}
\begin{cases}
&x_{i,t}=\beta_{i1}x_{1,t-1}+u_{i,t},\:\:i=1,2 \\
& u_{i,t} = \sigma_{i,t}\epsilon_{i,t}\\
&\sigma_{i,t}^2=\gamma_{i,0}+\gamma_{i,1}\sigma_{i,t-1}^2+\gamma_{i,2}u_{i,t-1}^2\\
&\epsilon_{i,t}\sim N(0,1),\:\:\:i.i.d.\:r.v.\:\forall i=1,2,
\end{cases}
\ee
then using both the estimated parameters and the filtered series to find at each time the Value at Risk $VaR_{5\%}$, \ie the $5\%$ left quantile corresponding to $-1.64\:\hat{\sigma}_{i,t}$ under the hypothesis of Gaussianity. Hence, we can obtain the binary time series of extreme events according to the condition $x_{i,t}<-1.64\:\hat{\sigma}_{i,t}$ which identifies the (left) tail events for the the time series $x_i$.\footnote{Here, we are considering the left tail of the distribution by finding the Value at Risk for the $5\%$ probability level. We can consider equivalently the right tail by using, \eg, the $95\%$ probability level.}

We apply the two tests for Granger causality in tail to data generated in the three different cases, namely NULL, ALTER1, and ALTER2. Under NULL, there is no Granger causality in tail from $x_2$ to $x_1$, thus we can study the size of each method, \ie the False Positive Rate (FPR), by testing under the null hypothesis that $x_2$ `does not Granger-cause in tail' $x_1$. On the other hand, there exists Granger causality in tail from $x_2$ to $x_1$ under both ALTER1 and ALTER2, but for different underlying effects. Under ALTER1, $x_2$ triggers an extreme event for $x_1$ by `moving' the conditional mean of the distribution of $x_1$. Under ALTER2, $x_2$ modifies the conditional variance of the distribution of $x_1$, thus increasing, \eg, the probability of a tail event. In both cases, we can study the power of each test of Granger causality in tail by finding the rate of rejections of the null hypothesis that $x_2$ `does not Granger-cause in tail' $x_1$. The results for both tests are shown in Table \ref{TablenullDGPbidarp}. In comparing the finite size performances, the method by \cite{hongetal} slightly outperforms ours. 

{\footnotesize
\begin{center}
\begin{table}
{\begin{flushleft} Size at the $5\%$ significance level of the test by Hong et al. or the LR test.\end{flushleft}}
 \begin{tabular}{||c c c c c c ||} 
 \hline
\thead{GARCH-type m.\\NULL} & \thead{FPR\\ Hong et al.\\ $M=5$} & \thead{FPR\\ Hong et al.\\ $M=10$} & \thead{FPR\\ Hong et al.\\ $M=15$} &\thead{FPR\\ Hong et al.\\ $M=20$} & \thead{FPR\\ LR test \\ $$}\\ [0.5ex] 
 \hline\hline
$T=500$ & 0.07 & 0.06 & 0.08  & 0.07 & 0.02\\ 
 \hline
$T=1000$ & 0.05 & 0.06 & 0.07 & 0.06 & 0.02\\ 
 \hline
$T=2000$ & 0.05 & 0.07 & 0.06  & 0.06 & 0.02\\ 
 \hline
 $T=5000$ & 0.08 & 0.06 & 0.07  & 0.06 & 0.03\\ 
 \hline
$T=10000$ & 0.07 & 0.09 & 0.07  & 0.08 & 0.02\\ [1ex] 
 \hline
\end{tabular}

\vspace{0.2cm}
{\begin{flushleft} Power at the $5\%$ significance level of the test by Hong et al. or the LR test.\end{flushleft}}
\begin{tabular}{||c c c c c c ||} 
 \hline
\thead{GARCH-type m.\\ALTER1} & \thead{TPR\\ Hong et al.\\ $M=5$} & \thead{TPR\\ Hong et al.\\ $M=10$} & \thead{TPR\\ Hong et al.\\ $M=15$} &\thead{TPR\\ Hong et al.\\ $M=20$} & \thead{TPR\\ LR test \\ $$}\\ [0.5ex] 
 \hline\hline
$T=500$ & 0.30 & 0.31 & 0.34  & 0.25 & 0.23\\ 
 \hline
$T=1000$ & 0.56 & 0.48 & 0.46 & 0.40 & 0.37\\ 
 \hline
$T=2000$ & 0.72 & 0.70 & 0.63  & 0.42 & 0.57\\ 
 \hline
 $T=5000$ & 0.98 & 0.95 & 0.94  & 0.92 & 0.93\\ 
 \hline
$T=10000$ & 1.00 & 1.00 & 1.00  & 1.00 & 0.99\\ [1ex] 
 \hline
\end{tabular}

\vspace{0.2cm}
{\begin{flushleft} Power at the $5\%$ significance level of the test by Hong et al. or the LR test.\end{flushleft}}
\begin{tabular}{||c c c c c c ||} 
 \hline
\thead{GARCH-type m.\\ALTER2} & \thead{TPR\\ Hong et al.\\ $M=5$} & \thead{TPR\\ Hong et al.\\ $M=10$} & \thead{TPR\\ Hong et al.\\ $M=15$} &\thead{TPR\\ Hong et al.\\ $M=20$} & \thead{TPR\\ LR test\\ $$}\\ [0.5ex] 
 \hline\hline
$T=500$ & 0.37 & 0.39 & 0.40  & 0.36 & 0.19\\ 
 \hline
$T=1000$ & 0.51 & 0.53 & 0.51 & 0.46 & 0.27\\ 
 \hline
$T=2000$ & 0.70 & 0.78 & 0.70  & 0.75 & 0.44\\ 
 \hline
 $T=5000$ & 0.97 & 0.98 & 0.97  & 0.97 & 0.81\\ 
 \hline
$T=10000$ & 1.00 & 1.00 & 1.00  & 1.00 & 0.98\\ [1ex] 
 \hline
\end{tabular}

\caption{FPR and TPR of the two statistical tests (Hong et al. test for different values of M and the LR test with the statistic $\Lambda$ (\ref{Lstatistics}) based on the VDAR(p) model) with data generated by the GARCH-type model as in \cite{hongetal} under the three different cases explained in the text (Upper: NULL, Middle: ALTER1, Bottom: ALTER2). $T$ indicates the sample size of each time series. The Danielsson kernel used for the Hong et al. test is with $M$ as indicated in the Tables. Each value is the average over a sample of $500$ seeds.}
\label{TablenullDGPbidarp}
\end{table}
\end{center}}

\begin{figure}
\includegraphics[width=0.49\textwidth]{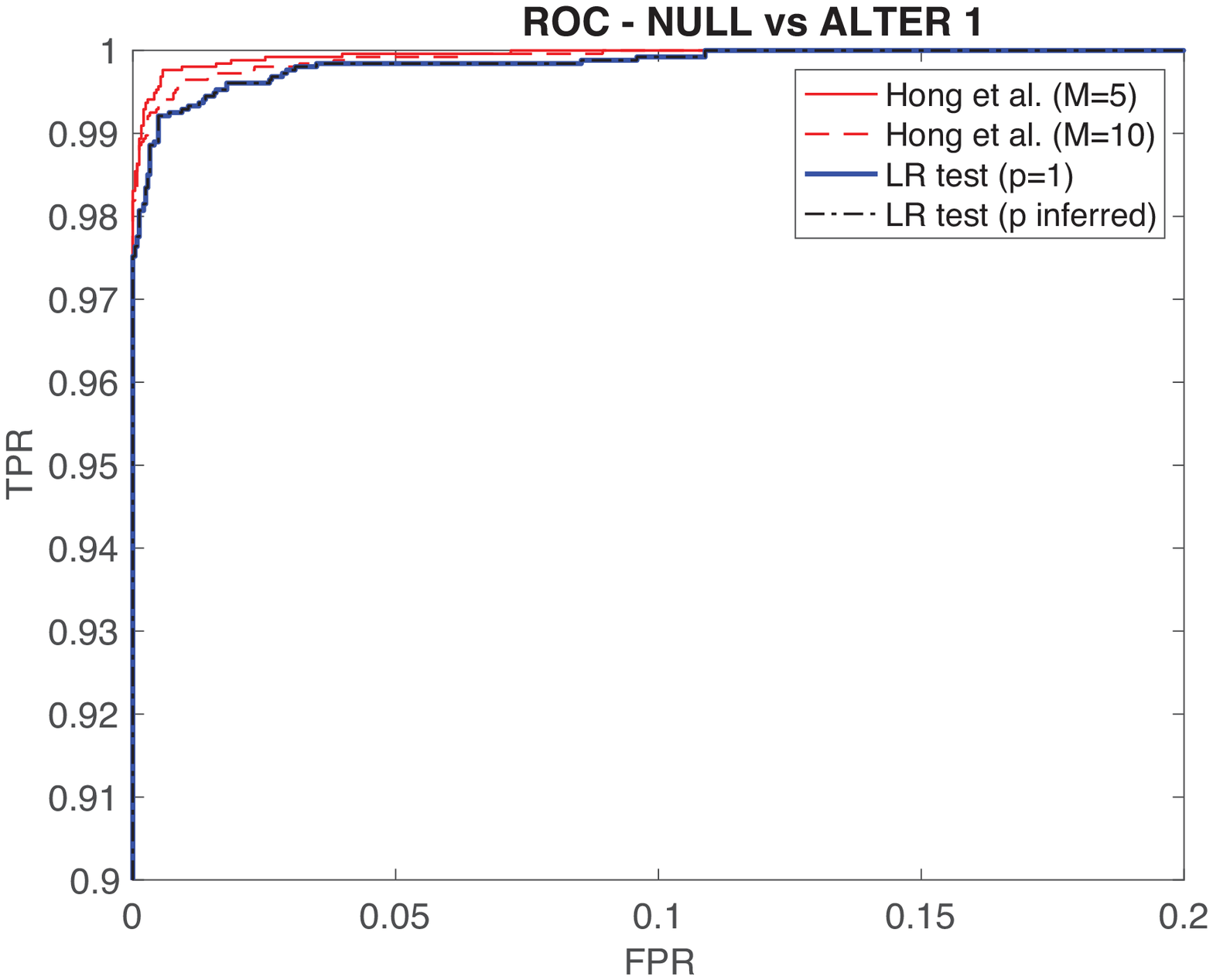}
\includegraphics[width=0.49\textwidth]{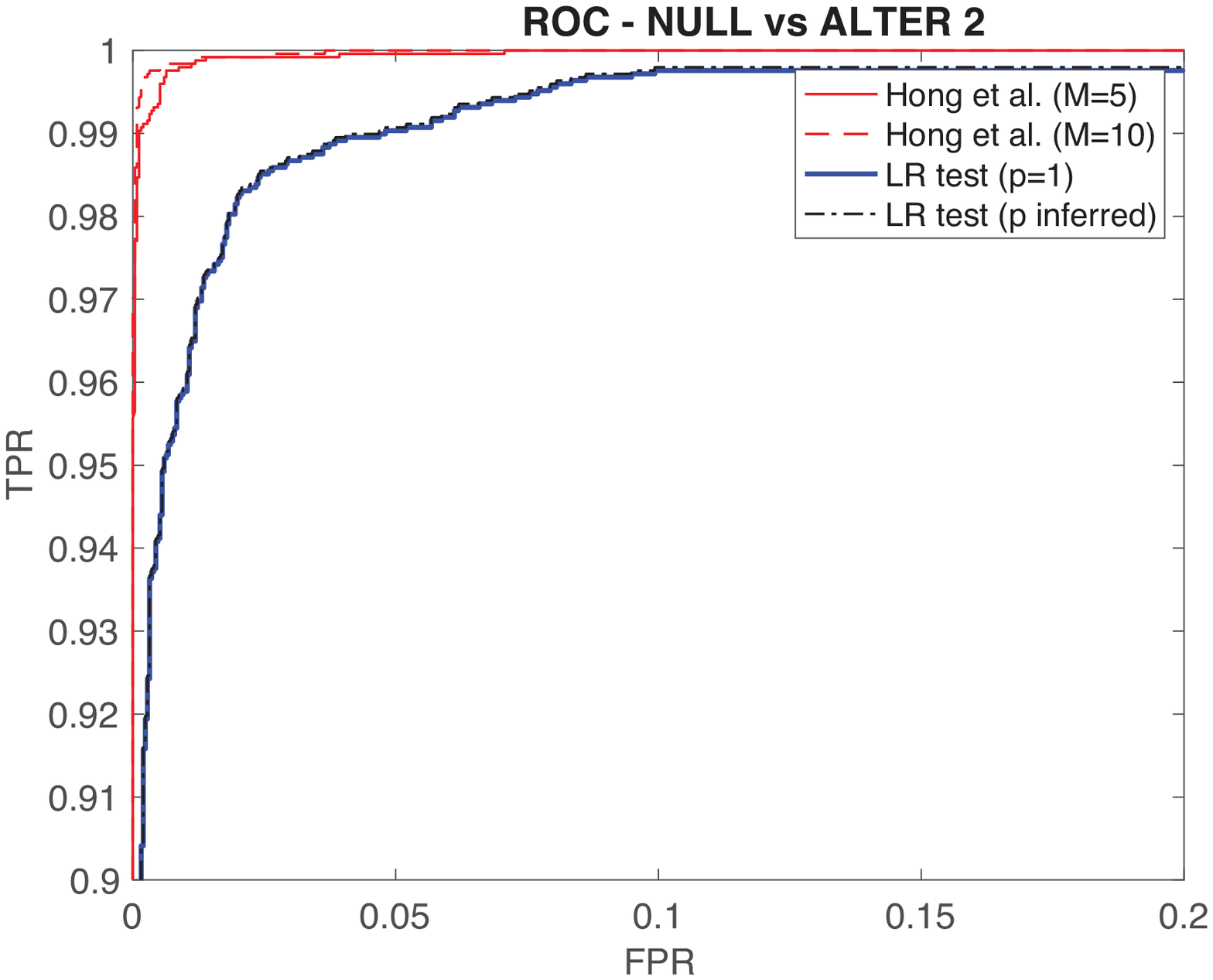}
\caption{Receiving Operating Characteristic (ROC) curves built with the p-values associated with the test statistics of either the Likelihood-Ratio (LR) method (\ref{Lstatistics}) and the kernel-based non-parametric approach by \cite{hongetal}, used as binary classifiers depending on some threshold value, \ie the test significance level. We use the GARCH-type model with $T=10000$ as Data Generating Process (DGP), under the null hypothesis NULL or under the alternative hypothesis ALTER1 (left) or ALTER2 (right), with one half probability over a sample of $5000$ simulations. The model parameters are as in the main text. In implementing the LR test, the blue curve is obtained by imposing the order $p=1$ of the autoregressive process, whereas the black dotted curve is obtained by optimally selecting the order with the Bayesian Information Criterion. The method by Hong et al. is implemented with the Daniell kernel with a $M=5$ (red solid line) or $M=10$ (red dotted line).}
\label{fig_comp2}
\end{figure}

As in the previous section, for a fair comparison we build the ROC curves associated with the two methods, see Figure \ref{fig_comp2}.  We simulated data according to the GARCH-type model, either without (NULL case) and with (ALTER1 case in the left panel, ALTER2 case in the right panel) causality, and computed the FPR and TPR obtained by the two methods. We used a large sample size $T=10000$. For both mechanisms of causality, ALTER1 and ALTER2, our method displays a very good performance as testified by the area under the (ROC) curve really close to one, a signal of high prediction power.  Nevertheless, the test by Hong et al. is more powerful in detecting causality, especially under ALTER2. Therefore, when data are generated by the considered GARCH-type models and in the pairwise scenario, the non-parametric approach by \cite{hongetal} performs slightly better than the proposed LR test. 

\begin{figure}
\includegraphics[width=0.49\textwidth]{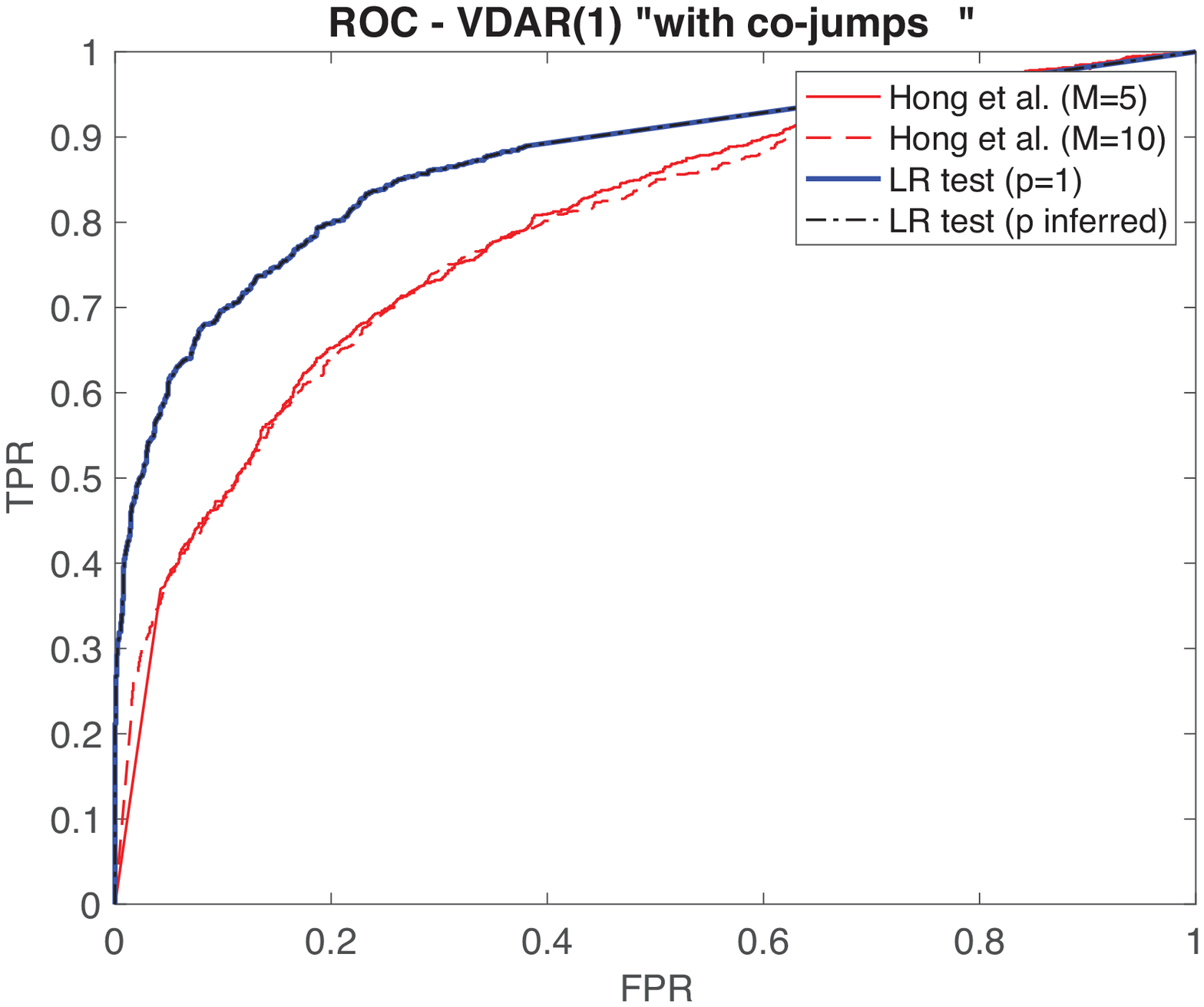}
\includegraphics[width=0.49\textwidth]{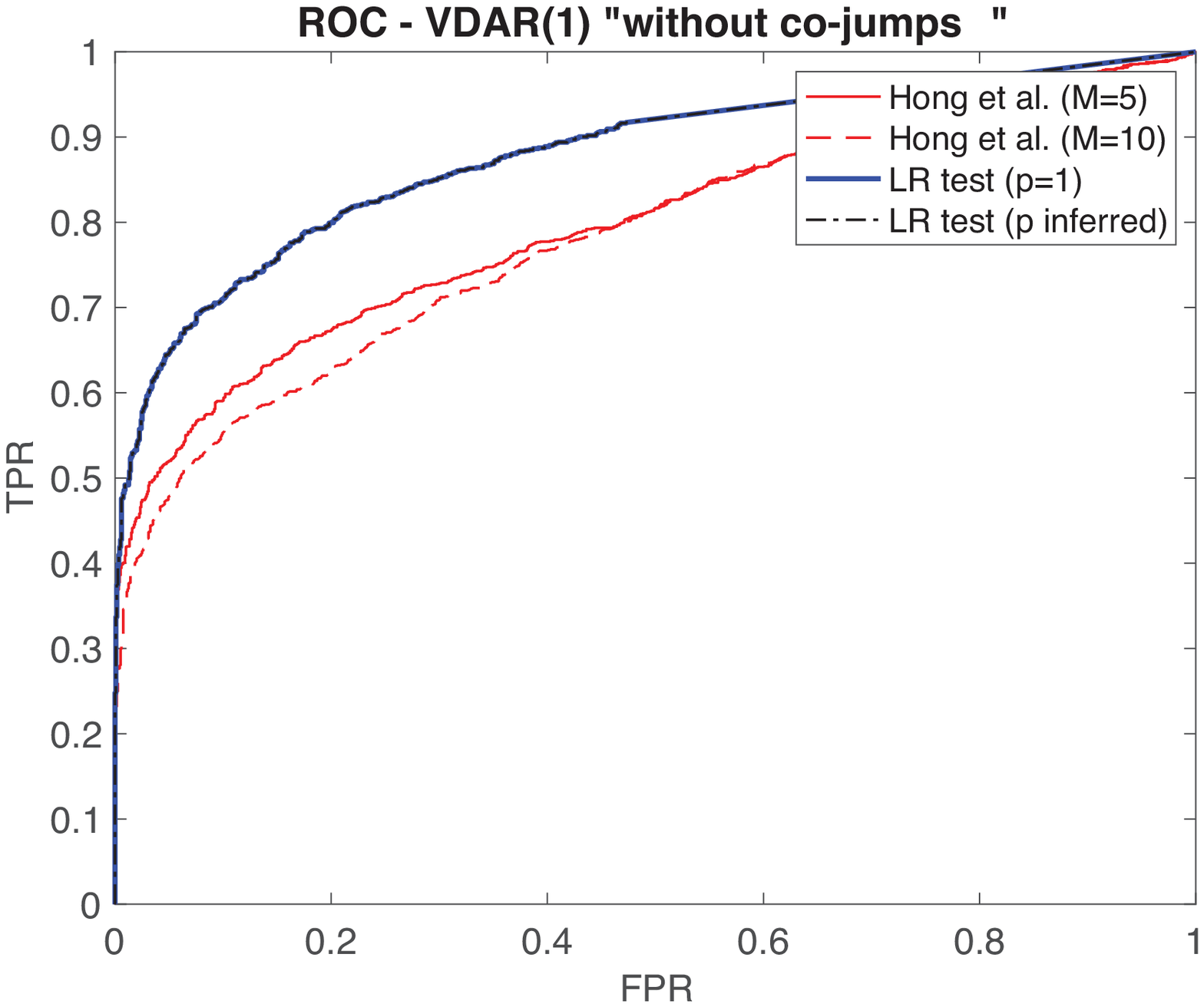}
\caption{Receiving Operating Characteristic (ROC) curves built for both the Likelihood-Ratio (LR) test (\ref{Lstatistics}) and the kernel-based non-parametric approach by \cite{hongetal}. Data are generated by the VDAR(1) model (\ref{bidarp}) with $p=1$ and sample size $T=10000$, but Bernoulli marginals with a dependence structure described by a Gaussian copula having zero mean and correlation parameter equal to $0.75$ (left) or $-0.75$ (right). The model parameters are: $\nu_1=0.05$, $\nu_2=0$, $\chi_1,\chi_2=0.05$, $\lambda_2=0$, and $\lambda_1=0$ under the null $\mathbb{H}^0$, or $\lambda_1\sim \mathcal{U}(0,1)$ under the alternative hypothesis $\mathbb{H}^A$. In implementing the LR test, the blue curve is obtained by imposing the order $p=1$ of the autoregressive process, whereas the black dotted curve is obtained by optimally selecting the order with the Bayesian Information Criterion. The method by Hong et al. is implemented with the Daniell kernel with a $M=5$ (red solid line) or $M=10$ (red dotted line).}
\label{fig_comp3}
\end{figure}

%However, we must not conclude that in general, when the data generating process is not correctly specified, the non-parametric approach performs better. 
{However, there are cases when, even if the data generating process is not correctly specified, the parametric approach works better.} In fact, consider the bivariate VDAR(1) model (\ref{bidarp}) with $p=1$ as data generating process, but with a general dependence structure between the Bernoulli marginal distributions, \ie $Z_t^1$ and $Z_t^2$ in (\ref{bidarp}), described by a Gaussian copula.
A positive (instantaneous) correlation structure between extreme events of prices (usually referred as {\it jumps} in this context) is quite common in financial markets, see \eg an empirical study on synchronization of large price movements in the US stock market \cite{calcagnile}. Again, we compare the ROC curves built for the two methods of Granger causality in tail, see Figure \ref{fig_comp3}. In this case, the test by Hong et al. is quite sensitive to the presence of instantaneous ``co-jumps'', resulting in an increased rate of false rejections (left panel), whereas our method assures a high performance in assessing causality relations at the unit time lag. For completeness, we report also the result for negative correlation between extreme events (\ie very low probability for the co-occurrence) in the right panel. Also in this case, our method outperforms the Hong et al. test. Therefore, the bivariate VDAR(1) with non-independent marginals is an example in which our parametric method works better than the non-parametric one even though the data generating process does not coincide with the one assumed by the method. 

\subsection{VDAR(1) as data generating process}
\label{sec:Vdar}
\begin{figure}
\includegraphics[width=0.99\textwidth]{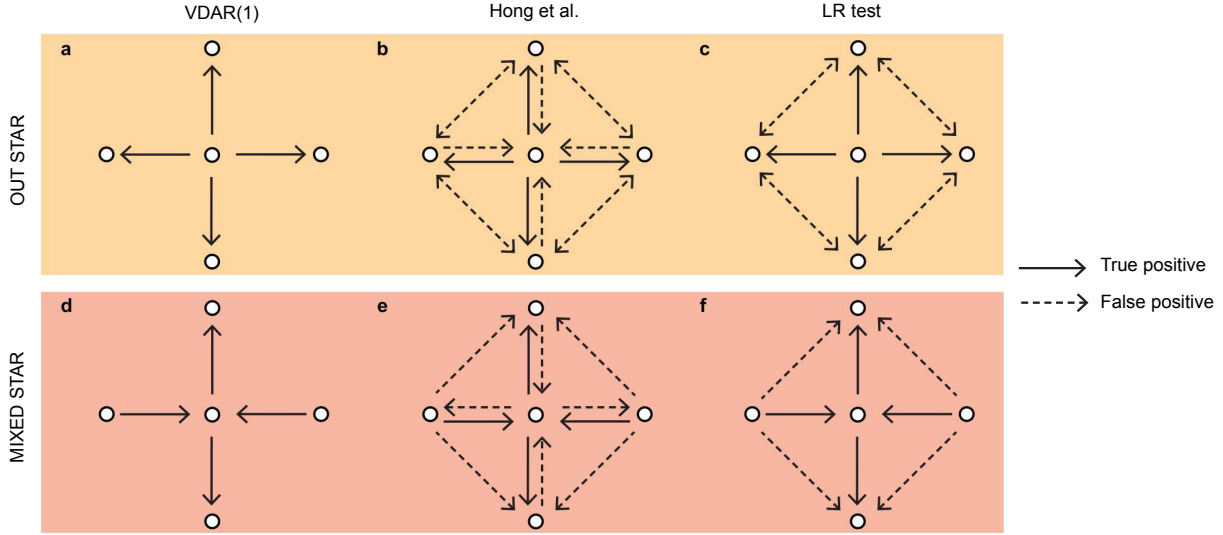}
\caption{Pictorial representation of both {\it out} and {\it mixed stars}, as explained in the main text, (a and d) together with the validated networks of interactions by means of the two methods of Granger causality in tail, namely Hong et al. (b and e) and LR test (c and f).}
\label{figNetEffects}
\end{figure}

Let us consider now the VDAR(1) model (\ref{vdar1}) as data generating process. We consider a system of $N=40$ variables with star-shaped network of causal interactions, meaning that the off-diagonal interaction terms $\lambda_{ij}$ are present between a central node (representing one of the variables) and each of the other $N-1$. The {\it extreme} states of these variables are described by binary time series evolving according to (\ref{vdar1}) with causal interactions captured by the matrix $\bm{\lambda}$ parameterized as
\be
\bm{\lambda} =
   \begin{pmatrix} 
   \lambda_{self}^{1} & \lambda_{in}\bm{u} '   \\
   \lambda_{out}\bm{v} & \diag(\lambda_{self}^i) &   \\
   \end{pmatrix},
\ee 
where $\diag(\lambda_{self}^i)$ is a $N-1\times N-1$ diagonal matrix having entries $\lambda_{self}^i$, sorted according to the node index $i=2,...,N$. The first node is the center of the star. It interacts with itself (at the past time lag) with probability $\lambda_{self}^1$, and with each of the other nodes with probability $\lambda_{out}$ or $\lambda_{in}$, respectively for outward or inward interactions. Any other node $i$ interacts with itself (at the past time lag) with probability $\lambda_{self}^i$, $i=2,...,N$, and with the central node. In particular, we consider two cases: (i) the {\it out} star where the $N-1$ nodes are `Granger-caused' by the first node, obtained setting $\lambda_{in}=0$, $\lambda_{out}=1/2$, and $\bm{v}\equiv \bm{1}_{N-1}$ (the vector of $N-1$ ones). In this case, $\lambda_{self}^{1}=1$ and $\lambda_{self}^i=1/2$ $\forall i=2,...,N$ (as required by normalization); (ii) the {\it mixed} star, where a subset of the nodes is `Granger-caused' by the first node, which, in turn, is `Granger-caused' by the complementary set, obtained by setting $\lambda_{out}=1/2$, $\bm{u}$ and $\bm{v}$ random vectors of ones and zeros such that $\bm{u}+\bm{v}=\bm{1}_{N-1}$\footnote{$\bm{u}$ is a random vector having each entry equal to either one or zero, depending on the realization of a Bernoulli random variable with one half success probability. The vector $\bm{v}$ is complementary to $\bm{u}$, such that $\bm{u}+\bm{v}=\bm{1}_{N-1}$.}, and $\lambda_{in} = 1/\left(1+\sum_{j=1}^{N-1} u_j\right)$. In this case, $\lambda_{self}^1=1/\left(1+\sum_{j=1}^{N-1} u_j\right)$ while $\lambda_{self}^i$ with $i\neq 1$ is $1/2$ if $i$ is `Granger-caused', $1$ otherwise (again, as required by normalization). In both cases, we set $\nu_i=1/2$ and $\chi_i=\chi$ $\forall i=1,...,N$, with varying $\chi\in(0,1/2]$. A pictorial representation of the two cases is shown in panels a and d of Figure \ref{figNetEffects}.

\begin{figure}
\includegraphics[width=0.99\textwidth]{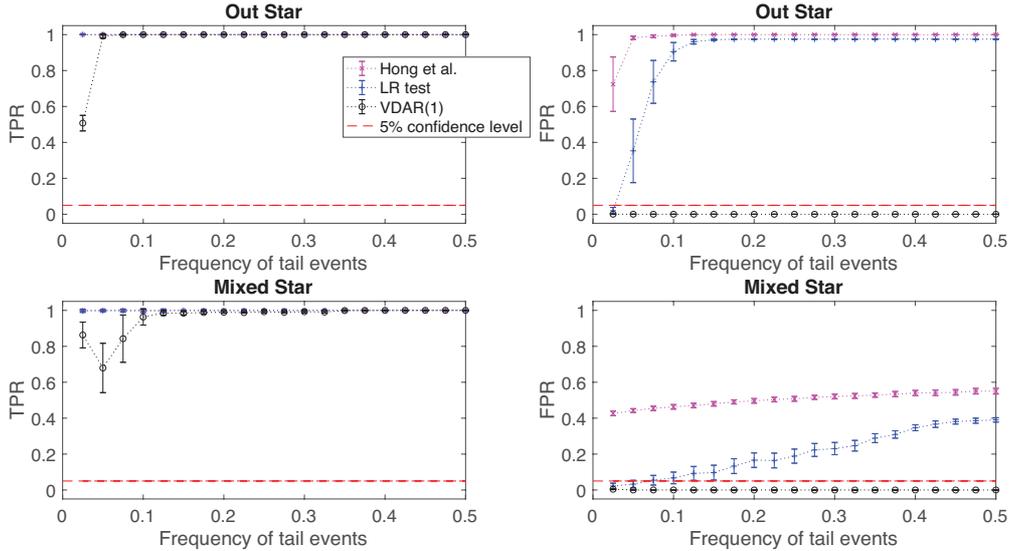}
\caption{True Positive Rate (TPR) and False Positive Rate (FPR) as a function of the frequency of tail events (or, equivalently, $\chi_i=\chi\in(0,1/2),\:\forall i=1,...,N$ in (\ref{vdar1})) for the three methods: the test by Hong et al., the LR test with statistic $\Lambda$ (\ref{Lstatistics}), and the statistically validation of the interaction terms of the VDAR(1) process by means of Decimation. Data are generated by the VDAR(1) model as described in the main text. Top panels refer to the case of {\it out} star, while bottom panels to {\it mixed star}. The significance level is $5\%$, but corrected with the False Discovery Rate (FDR) method to take into account the effect of multiple testing comparison. Each value and the corresponding error bar are the mean and the standard deviation, respectively, over a sample of 100 simulation.}
\label{figStars}
\end{figure}

Hence, a node `Granger-causes' another node if a directional link exists in the star. Thus, once the data are generated by the VDAR(1) model, we can obtain both the power and the size of the statistical tests of Granger causality in tail by considering how each method is able to reconstruct on average the network of interactions. Notice that both the statistical method by \cite{hongetal} and the LR test based on the statistic $\Lambda$ (\ref{Lstatistics}), being {\it pairwise} approaches, detect causal interactions in the network by considering sequentially couples of nodes, whereas the statistical validation of the interactions in the VDAR(1) process by means of Decimation is an effectively {\it multivariate} approach. 

The TPR and FPR for the three methods in each of the two cases, namely the {\it out} and the {\it mixed} stars, are shown in Figure \ref{figStars}. The rejection rates are plotted for varying frequency of tail events, controlled by the parameter $\chi$. In the left panels, we notice that all three methods are powerful in rejecting in both cases the null hypothesis when a causal link is present in the stars. However, the right panels show that the rate of false rejections is quite high when we adopt the method by Hong et al. or the LR test. When considering the {\it out} star, the FPR associated with the method by Hong et al. converges quickly to one as the frequency of tail events increases, \ie the null hypothesis is always rejected for any possible couple of non-interacting nodes. The approach based on the LR test outperforms slightly the method by Hong et al., nevertheless it displays a very high FPR. When considering instead the {\it mixed} star, the rate of false rejections of both tests is significantly larger than the significance level of $5\%$ (with the exception of the LR test in the case of really low frequency of tail events), and, again, the approach based on the test statistics $\Lambda$ (\ref{Lstatistics}) outperforms the method by Hong et al. Both the pairwise tests, therefore, have a non-satisfactory performance in presence of network effects. On the contrary, the {\it multivariate} approach based on the statistical validation of the interaction terms in the VDAR(1) process by means of Decimation shows a False Positive Rate below the threshold of the significance level, very close to zero, for both the {\it out} and the {\it mixed} stars. 

The very high rate of false rejections associated with the pairwise causality analysis is the combination of two effects: first, a misspecification of the {\it information set} and, second, non-zero autocorrelation. When considering the {\it out} star, the central node `Granger-causes' all the others, but false positives are detected in both directions between non-interacting nodes when the information on the central node is not considered (compare panels a and c of Figure \ref{figNetEffects}). This happens both for the LR test based on the statistic $\Lambda$ (\ref{Lstatistics}) and for the Hong et al. test. Moreover, when the binary time series have non-zero autocorrelation, the test by Hong et al. mistakenly detects an additional causal interaction from the outer nodes to the central node (panel b of Figure \ref{figNetEffects}). In this case, the validated network of interactions becomes, on average, the complete graph, \ie a graph with all possible links, since both the TPR and the FPR converge to one. Similar spurious causality arises for the {\it mixed} star. When $A$ causes $B$, and $B$ causes $C$, a spurious causality can be detected from $A$ to $C$, if $B$ is not considered. This causes spurious links in the network reconstructed by the LR test (compare panels d and f of Figure \ref{figNetEffects}) and also by the Hong et al. test. The latter also detects spurious links in presence of non-zero autocorrelation, see panel e. The multivariate approach is not prone to any of these mistakes, since it reconstructs correctly the networks of interactions shown in panels a and d.

%%% NEW
%%%%%%%%%%%%%%%%
\section{Empirical application on the US Stock Exchange market}\label{sec:application}
To illustrate our novel method for testing Granger causality in tail, we now consider an application to high frequency data of a portfolio of financial stocks belonging to the Russell 3000 Index, traded in the US equity markets (mostly NYSE and NASDAQ). 
%Data are provided by \href{http://www.kibot.com}{Kibot}. 
An empirical characterization of such high frequency data can be found in \cite{calcagnile}. Here, we adopt the same filtering procedure introduced in \cite{bormetti} to obtain the time series of the extreme events of stock returns, then we apply our method to detect the causal relations between stocks.

\subsection{Data filtering and extreme events of stock returns} We consider $39$ highly liquid stocks traded in the US markets from 2000 to 2012. We use one-minute closing price data during the regular US trading session, \ie from 9:30 am to 4:00 pm. Hence, we define the return at the one minute time scale as $\tilde{r}_t^i\equiv\log P_t^i/P_{t-1}^i$ where $P_t^i$ is the price of stock $i$ at time $t$, with $i=1,...,39$ and the $t=1,...,T$ where $T$ is about $98000$ per year (this number depends on the effective number of trading days within the year).

Intraday returns are first filtered because of the presence of the well-known U-shape pattern for volatility within a trading day, \ie prices exhibit typically larger movements at the beginning and at the end of the day. The {\it raw} return $\tilde{r}_{d,t}^i$ of generic stock $i$ at day $d$ and intraday time $t$ is rescaled by a factor $u_t^i$,
\be\nonumber
r_{d,t}^i=\frac{\tilde{r}_{d,t}^i}{u_t^i},
\ee
where 
\be\nonumber
u_{t}^i=\frac{1}{N_{days}}\sum_{d'}\frac{\vert\tilde{r}_{d',t}^i\vert}{s_{d'}^i},
\ee
with $N_{days}$ indicating the number of days in the sample and $s_{d'}^i$ the standard deviation of absolute intraday returns of day $d'$. Rescaled returns $\{r_t^i\}_{t=1,...,T}^{i=1,...,39}$ no longer possess any daily regularities.%\footnote{\textcolor{blue}{In principle, the described standard procedure to filter out the intraday pattern could be problematic since it uses future returns to rescale the return at the current time, thus introducing future information in a testing procedure which is based on forecasting. Nevertheless, the intraday seasonality pattern tends to be quite constant in our sample, independently from the specific days or time periods. Moreover, the whole information is aggregated into a single value per time of day, hence returns are rescaled by the same factor, depending on the time of the day. Thus, all the past and future information is aggregated to a single value (per time of the day). Thus, it can be expected that the filtering procedure does not affect importantly the output. This can be verified, for example, by using rescaling factors exploiting only the past information of the return dynamics.}}
\footnote{{In principle, the described procedure to filter out the intraday pattern could be problematic since it uses future returns to rescale the return at the current time, thus introducing future information in a testing procedure which is based on forecasting. Nevertheless, the intraday seasonality pattern tends to be quite constant in our sample. For this reason it can be expected that the filtering procedure does not significantly affect the output, as we verified by using rescaling factors estimated using only past information. The latter procedure,  however, requires removing from the analysis a significant amount of data, while we prefer to present here results for the whole sample.}}

To estimate the spot (\ie instantaneous) volatility $\sigma_t^i$, we use the method of {\it realized bipower variation}\footnote{{The method of realized bipower variation with threshold estimation for volatility filtering is preferred to the estimation method based on realized variance, since it has been proved the existence of a bias in the presence of  jumps, see \cite{barndoff2}, by assuming the existence of an underlying continuous-time stochastic volatility process.}} by \cite{barndoff2} with {\it threshold} correction for the presence of jumps by \cite{corsiTBV} and using exponentially weighted moving averages of returns, see also \cite{bormetti}. {That is, volatility is estimated through returns which are not identified as jumps (\ie returns whose absolute value is not larger than a threshold value $\theta$ times the local volatility). Using exponentially weighted moving averages, the spot variance is estimated recursively as}
\be\nonumber
(\sigma_t^i)^2 = \mu_1^{-2}\alpha\vert r^i_{t''} \vert \vert r^i_{t'}\vert+(1-\alpha)(\sigma_{t-1}^i)^2
\ee
{with $\mu_1=\sqrt{2/\pi}$, $\alpha=2/61$, and where $t''$ and $t'$ are such that $t''<t'\leq t-1$, $\frac{\vert r^i_{t''}\vert}{\sigma^i_{t''}}\leq\theta$, $\frac{\vert r^i_{t'}\vert}{\sigma^i_{t'}}\leq\theta$, and $\frac{\vert r^i_{\tau}\vert}{\sigma^i_{\tau}}>\theta$ for each $t''<\tau<t'$ and for each $t'<\tau<t$. The value of the parameter of the exponentially weighted moving average $\alpha$ corresponds to an effective time window of about $20$ minutes.}

Finally, we say that an extreme return occurs for stock $i$ at time $t$ when
\be\label{eqjump}
\frac{r_t^i}{\sigma_t^i}<-\theta.
\ee

{Notice that, after removing intraday seasonality and volatility patterns, returns are approximately Gaussian distributed \cite{cont}. Then the parameter $\theta$ determines the quantile of the return distribution which is considered to be a tail, \ie $-\theta \times\sigma_t^i$. We use $\theta=4$ in the present analysis, corresponding to a probability of a tail event equal to about $0.1\%$. This choice is motivated by a sensitivity analysis of Granger causality networks built for different values of $\theta$, where we have found that the outputs of the causality analysis are qualitatively similar in a range around $\theta=4$.}

By using condition (\ref{eqjump}), we build the binary time series $\{X_t^i\}_{t=1,\dots,T}^{i=1,\dots,39}$ of extreme events for the US stocks.

\subsection{Causal relations in the US stock exchange market}
We now consider the directed network of causal interactions between extreme events of the underlying stock return dynamics. Again, we stress that the output of the causality analysis depends largely on the adopted method, thus the importance of understanding strengths and weaknesses of each statistical test of Granger causality in tail. 

In building the Granger causality network, we can adopt either a multiple hypothesis testing approach based on pairwise causality analysis or the multivariate approach based on the statistical validation of off-diagonal couplings. In the first case, given $N$ time series of extreme events and considering either the novel LR test or the one by Hong et al., we construct the network by applying the Granger causality test to all the possible $N(N-1)$ pairs. Then, assume we aim to obtain some overall significance level for the multiple testing. Because of $N(N-1)$ simultaneous tests, a correction to the significance level of each single test needs to be considered to take into account the increased chance of rare events and therefore, the increased probability of false positive (i.e. rejections) \cite{tumminello2011statistically}. In the present analysis, the overall significance level is set to $5\%$ and the False Discovery Rate (FDR) method \cite{ben} is applied to correct the p-value of each single test. In the second case, adopting the multivariate approach based on the VDAR(1) model, the network of causal interactions is validated by means of Decimation, thus the overall significance level is implicitly defined by the validation procedure. Finally, we use the three different methods to build the corresponding Granger causality networks with the data of (left) extreme events of $39$ US stocks filtered as above, and considering one year at the time from $2000$ to $2012$.

\begin{figure}
\includegraphics[width=0.99\textwidth]{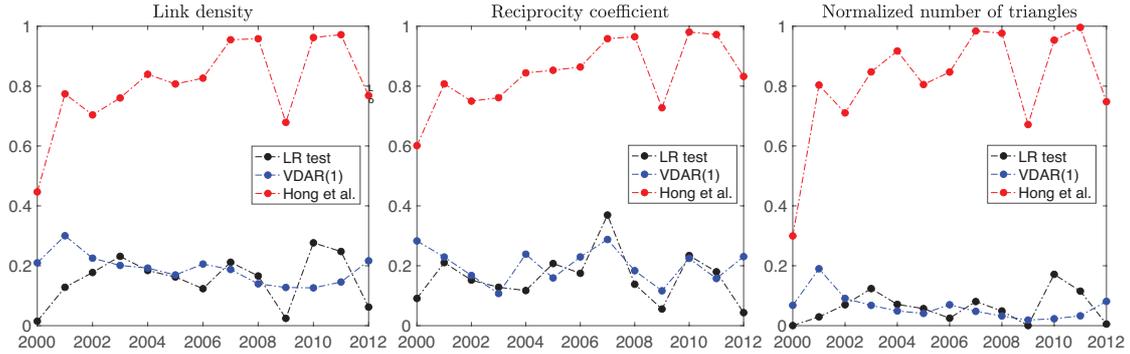}
\caption{Link density (left), reciprocity coefficient (middle), and normalized number of closed triangles (right) of the Granger causality networks obtained with the three different methods, \ie the LR test with test statistic (\ref{Lstatistics}) (black), the test by \cite{hongetal} (red), and Decimation applied to the VDAR(1) model (blue).}
\label{figGCnet39stocks}
\end{figure}
\begin{figure}
\includegraphics[width=0.99\textwidth]{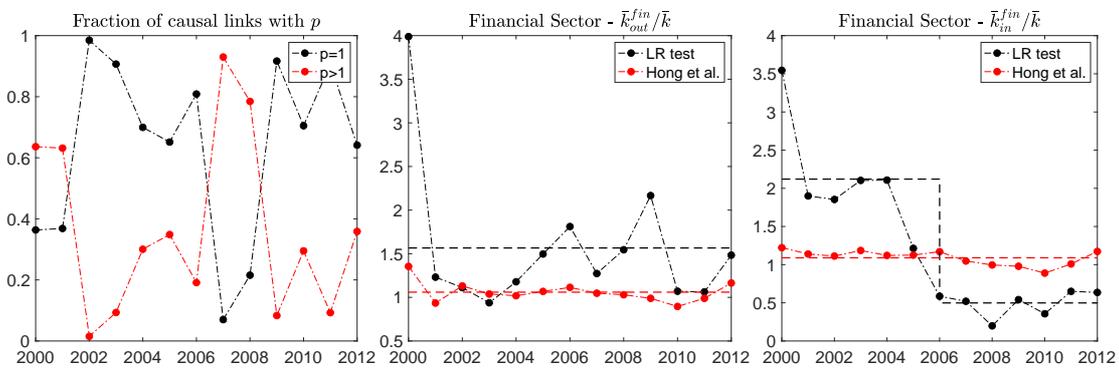}
\caption{Fraction of causal links validated by the LR test and described by the VDAR(p) model with $p$ (as in the legend) optimally selected by the Bayesian Information Criterion (left). Mean node degree for both outgoing (middle) and incoming (right) causal links of the US stocks belonging to the Financial Sector, namely BAC, C, AXP, and WFC, \ie $\bar{k}_{out}^{fin}$ and $\bar{k}_{in}^{fin}$ respectively, are rescaled by the mean overall degree $\bar{k}$ of the corresponding Granger causality network. Dotted lines represent the mean over the period, for the Granger causality networks obtained according to the LR test (black) or Hong et al. (red). In the right panel, the mean value of the rescaled in-degree for the LR test is computed over two disjoint time windows, before and after $2006$.}
\label{figGCpk}
\end{figure}

Figure \ref{figGCnet39stocks} shows some characteristics of the different causality networks. First, the number of validated causal interactions differs significantly if we use the methods based on discrete autoregressive processes or the one proposed by Hong et al., the latter describing an almost complete graph, see the left panel of Figure \ref{figGCnet39stocks}. By using the Jaccard index\footnote{The Jaccard similarity coefficient between two sample sets is defined as the size of their intersection divided by the size of their union (of the links in the two networks).} to compare two causality networks, we measure a value between $0.4$ and $0.5$ for the Granger networks obtained with both the pairwise VDAR and the multivariate approach for almost all periods, signalling many common causal links between the two networks. On the contrary, we measure a much smaller value (between $0.1$ and $0.2$) when comparing the Granger network obtained with the test of Hong et al. with the others. In the previous section, we noticed how spurious effects may be present in the multivariate case if we adopt a pairwise approach. In particular, the detection of causality in both directions or in triangular loops may be due to non-zero autocorrelation and misspecification of the information set. Possible metrics capturing such network effects are the {\it reciprocity coefficient}\footnote{The reciprocity coefficient is defined as the ratio of the number of links pointing in both directions to the total number of links.} measuring the likelihood of nodes to be mutually linked, and the number of {\it closed triangles}\footnote{A normalized measure for the number of closed triangles in a network can be defined as the ratio of the number of subgraphs of three nodes (\ie triplets) which are connected each other independently from link directions, with the number of all possible triplets, both open and closed.}. The middle panel and the right panel of Figure \ref{figGCnet39stocks} show the two network metrics for the three Granger networks and, again, we can notice how reciprocated casual links and triangular interactions are largely over-expressed in the causality network obtained with the method of Hong et al., thus suggesting the presence of spurious detections. It is interesting to notice that during the first year of the US financial crisis of $2007-2008$ reciprocity of causal links in the network obtained with the LR test displays a significant increase with respect to the mean level of reciprocity during the whole considered period. 

An interesting empirical observation concerns how memory in causal links depends on the financial cycle. We notice that during turmoil periods for the US stock exchange markets, \ie the dot-com bubble of $2000-2001$ and the subprime mortgage crisis of $2007-2008$, %there %is a switching behavior for the Markovian dynamics of the causal links detected by the LR test. In particular,
non-Markovian effects become more important, as testified by the fraction of causal links better described by a VDAR(p) model with $p>1$ (optimally selected by the Bayesian Information Criterion), see the left panel of Figure \ref{figGCpk}. 

Finally, we consider the stocks belonging to the financial sector, \ie BAC, C, AXP, and WFC. The middle and right panel of Figure \ref{figGCpk} show the average value over this set of stocks of both the out-degree and the in-degree, \ie the mean number of outgoing and incoming causal links, respectively. Each value is rescaled by the average degree of the causality network. We compare the Granger causality network obtained by either our method or the one of Hong et al. We find that the two methods describe the subsystem of financial stocks quite differently: while, according to Hong et al., the financial sector is on average equivalent to any other one, the VDAR method suggests that the financial sector `Granger-causes' more than the others over all the considered period, since the ratio in the middle panel of Figure \ref{figGCpk} is above one. Moreover, there is a transition around $2005-2006$ in the average number of causal links pointing to the financial stocks, thus highlighting a scenario where the other sectors began to cause less the financial sector before the crisis, while, at the same time, financial stocks started to cause more the other sectors. 
%This simple example shows how \textcolor{blue}{our novel method is able to reveal some patterns which, on the contrary, may be hidden by using other approaches more sensitive to auto-correlations, instantaneous correlations (resulting in ``co-jumps''), and/or feedback effects between financial time series of extreme events.}
{This example application shows how our novel method reveals causal patterns that go undetected using other approaches, thanks to its proper accounting of both auto-correlations, instantaneous correlations ({\ie ``co-jumps''}), and/or network effects between financial time series of extreme events.}
%different conclusions on the role of a stock or sector in the causal networks can be drawn depending on the adopted method.

\section{Conclusions}\label{sec:conclusion}
Based on the definition of Granger causality in tail, we propose a novel statistical approach in the original spirit of Granger, to test whether the information on extreme events of one time series is statistically significant in forecasting extreme events of a second time series, by introducing the multivariate generalization of the {\it discrete autoregressive} process, namely VDAR(p). We devise a method based on the Likelihood-Ratio test statistic which is able to detect causal interactions between two time series, while also correctly identifying the time scale of the causal interaction. Then, to overcome the limit of the pairwise causality analysis resulting in spurious detections because of neglected variables, we also propose a statistical method for the multivariate case with Markovian dynamics. Finally, we highlight the importance of disentangling true causalities from false detections, and we prove the dependence of the results from the used statistical method. To this end, we present also a comparative study with the current standard in literature represented by the test of \cite{hongetal}.

Simulation studies show that the proposed test of Granger causality in tail has both good power and good size in finite samples, against a large variety of data generating processes. We show numerically that the proposed method and the test of Hong et al. differ under some circumstances. In particular, the test by Hong et al. displays some sensitivity to auto-correlation of the time series of extreme events, resulting in spurious effect of two-way causality detection in the presence of unidirectional relations, a drawback which is solved by our method. Then, we prove numerically how network effects may result in spurious detections of triangular interactions due to the misspecification of the information set in the pairwise causality analysis, a second drawback which is solved by our multivariate approach. Furthermore, we highlight some signals which may be associated with false detections in networked systems, namely the high rate of either reciprocated causal links or triangular loops.

The empirical application to high frequency data of the US stock exchange points out how the output of the causality analysis depends significantly on the adopted statistical procedure. In particular, we find that the network of causal interactions is sparse and the memory of the causation process depends on the financial cycles, with significant non-Markovian effects during financial crises. A focus on the financial sector reveals that financial stocks `Granger-cause' more than the others, especially during turmoil periods, and started to be `Granger-caused' less before the financial crisis of $2007-2008$. On the contrary, the method by Hong et al. describes a network which is dense, without displaying any specific pattern for the financial sector, whose stocks are on average equivalent to the others.

\section*{Acknowledgements}

This project has received funding from the SESAR Joint Undertaking under the European Union's Horizon 2020 research and innovation programme under grant agreement No 783206. The opinions expressed herein reflect the authors' views only. Under no circumstances shall the SESAR Joint Undertaking be responsible for any use that may be made of the information contained herein.

\appendix

\section{Inference of Discrete Autoregressive Models}
\label{app:mle}

The Maximum Likelihood Estimators (MLE) of the VDAR(p) models (\ref{bidarp}) and (\ref{vdar1}) is as follows.
\subsection{MLE of bivariate VDAR(p)}
Assume to observe the binary time series $\{\bm{X},\bm{Y}\}\equiv\{X_t,Y_t\}_{t=1,...,T}$ and we aim to obtain the maximum likelihood estimator of the parameters $\nu_1,\lambda_1,\chi_1$, and $\bm{\gamma}_{11}\equiv\{\gamma_{11,k}\}_{k=1,...,p}$, $\bm{\gamma}_{12}\equiv\{\gamma_{12,k}\}_{k=1,...,p}$ (or, equivalently, $\nu_2,\lambda_2,\chi_2$, and $\bm{\gamma}_{21}\equiv\{\gamma_{21,k}\}_{k=1,...,p}$, $\bm{\gamma}_{22}\equiv\{\gamma_{22,k}\}_{k=1,...,p}$). The log-likelihood of observing the time series $\bm{X}$ given the information on $\bm{Y}$ according to the bivariate VDAR(p) model is
\be\label{logPbidarp}
\begin{split}
& \log \P(\bm{X}|\bm{Y},\nu_1,\lambda_1,\chi_1,\bm{\gamma}_{11},\bm{\gamma}_{12})=  \log\prod_{t=p+1}^T\P(X_{t}|X_{t-1},...,X_{t-p},Y_{t-1},...,Y_{t-p},\nu_1,\lambda_1,\chi_1,\bm{\gamma}_{11},\bm{\gamma}_{12}) = \\
& = \sum_{t=p+1}^T\log\left[\nu_1\left((1-\lambda_1)\sum_{k=1}^p\gamma_{11,k} \delta_{X_t,X_{t-k}}+\lambda_1\sum_{k=1}^p\gamma_{12,k}\delta_{X_t,Y_{t-k}}\right)+(1-\nu_1)(\chi_1)^{X_t}(1-\chi_1)^{1-X_t}\right],
\end{split}
\ee
by using the chain rule and conditioning on $\bm{Y}$ and the first $p$ observations $X_1,...,X_p$, where $\delta_{A,B}$ is the Kronecker delta taking value equal to one if $A=B$ and zero otherwise.

MLE of the parameters is obtained by maximizing the log-likelihood via gradient descendent methods \cite{tibi}, or finding the point estimate by solving iteratively the following system of equations
\be
\begin{cases}
\partial_{\phi}\log\P(\X|\X,\nu_1,\lambda_1,\chi_1,\bm{\gamma}_{11},\bm{\gamma}_{12}) &= 0,\:\:\:\:\forall \phi = \nu_1, \lambda_1,\chi_1,\\
\partial_{\gamma_{11,k}}\log\P(\X|\Y,\nu_1,\lambda_1,\chi_1,\bm{\gamma}_{11},\bm{\gamma}_{12}) &=0,\:\:\:\:\forall k=1,...,p-1,\\
\partial_{\gamma_{12,k}}\log\P(\X|\Y,\nu_1,\lambda_1,\chi_1,\bm{\gamma}_{11},\bm{\gamma}_{12}) &=0,\:\:\:\:\forall k=1,...,p-1,
\end{cases}
\ee
together with the conditions $\gamma_{11,p}=1-\gamma_{11,1}-...-\gamma_{11,p-1}$ and $\gamma_{12,p}=1-\gamma_{12,1}-...-\gamma_{12,p-1}$.

In both cases, the estimation algorithm needs a starting point. A random point in the parameter space can be used. However, we suggest to adopt the solution of the Yule-Walker equations for the VDAR(p) model.

\subsubsection{Yule-Walker equations for VDAR(p)}
By taking expectations of both sides of Equation (\ref{bidarp}), we obtain
\be\label{expbidarp}
\begin{cases}
\E(X_t) &= \nu_1((1-\lambda_1)\sum_{k=1}^p\gamma_{12,k}\E(X_{t-k})+\lambda_1\sum_{k=1}^p\gamma_{12,k}\E(Y_{t-k}))+(1-\nu_1)\chi_1\\
\E(Y_t) &= \nu_2(\lambda_2\sum_{k=1}^p\gamma_{21,k}\E(X_{t-k})+(1-\lambda_2)\sum_{k=1}^p\gamma_{22,k}\E(Y_{t-k}))+(1-\nu_2)\chi_2.
\end{cases}
\ee
This is formally equivalent to the expectation of a bivariate VAR(p) model \cite{tsay}
\be\label{yw1}
\E(\bm{Z}_t) = \bm{\phi}_0+\sum_{k=1}^p\Phi_{k}\E(\bm{Z}_{t-k})
\ee
with $\bm{\phi}_0\equiv\{\phi_{0,1},\phi_{0,2}\}$ and $\Phi_k$ as $2\times 2$ matrix $\forall k=1,...,p$, by using a suitable one-to-one mapping of parameters and with $\bm{Z}_t\equiv(X_t,Y_t)'$. 

It is also trivial to show the following identity
\be\label{ywk}
\E(\tilde{\bm{Z}}_t\tilde{\bm{Z}}_{t-k}') = \Phi_k\E(\tilde{\bm{Z}}_{t-1}\tilde{\bm{Z}}_{t-k}')\:\:\:\forall k=1,...,p,
\ee
where the tilde indicates the mean subtracted variable.

The Yule-Walker estimator of the parameters $\bm{\phi}_0,\Phi_{k},\:\:\forall k=1,...,p$ is a standard result of time series analysis, see \cite{tsay} for a reference, and can be obtained by solving the linear system (\ref{ywk}) for the entries of $\Phi_k\:\:\forall k=1,...,p$, then solving (\ref{yw1}) for $\bm{\phi}_0$. Finally, the estimated parameters of the bivariate VDAR(p) model are obtained by inverting the one-to-one mapping between (\ref{expbidarp}) and (\ref{yw1}).

\subsubsection{Optimal selection of $p$}
Assume to estimate the bivariate VDAR(p) model (\ref{bidarp}) for different orders $p$, thus asking for the value of $p$ describing better the data. The optimal $p$ can be selected by using the Bayesian Information Criterion (BIC) \cite{tibi}, as follows. The BIC statistic is defined as
\be\label{bic}
\mbox{BIC}^{(p)} = 2(2p+1) \log T - 2\log \P(\X,\Y|\hat{\Theta}^{(p)})
\ee
where $\hat{\Theta}^{(p)}$ is the MLE of the $2(2p+1)$ VDAR(p) parameters and $T$ is the sample size. Notice that
\be\label{lltot}
\log \P(\X,\Y|\hat{\Theta}^{(p)}) = \log \P(\X|\Y,\hat{\Theta}^{(p)})+ \log \P(\Y|\X,\hat{\Theta}^{(p)}),
\ee
where $\log \P(\X|\Y,\hat{\Theta}^{(p)})$ is as in (\ref{logPbidarp}) (and a similar relation holds for $\log \P(\Y|\X,\hat{\Theta}^{(p)})$), by using the chain rule. Finally, the optimal value of $p$ is the one associated with the lowest BIC.

\subsection{MLE of DAR(p)}
In implementing the Likelihood-Ratio test, the MLE of the DAR(p) model is required to obtain the test statistics (\ref{Lstatistics}). The maximum likelihood inference of the DAR(p) model (\ref{darp}) is similar to the case of VDAR(p), but without dependence on $\Y$, thus resulting in imposing $\lambda_1=0$ and $\gamma_{12,k}=0$ $\forall k=1,...,p$ in (\ref{logPbidarp}). Then, the inference process follows similarly as described before. The Yule-Walker estimator of the DAR(p) model is described in \cite{dar}. The optimal selection of the order $p$ is as before.

\subsection{MLE of multivariate VDAR(1) and Decimation of coupling parameters} Assume to observe the binary time series $\X\equiv\{X_t^i\}_{t=1,...,T}^{i=1,...,N}$ following the VDAR(1) process (\ref{vdar1}) and looking at the maximum likelihood estimator of model parameters $\bm{\nu}\equiv\{\nu_i\}_{i=1,...,N}$, $\bm{\lambda}\equiv\{\lambda_{ij}\}_{i,j=1,...,N}$ with conditions $\sum_{j=1}^N\lambda_{ij}=1$ $\forall i=1,...,N$, and $\bm{\chi}\equiv\{\chi_i\}_{i=1,...,N}$. The log-likelihood of observing the time series $\X$ is
\be\label{logPvdar1}
\begin{split}
&\ell(\X,\bm{\nu},\bm{\lambda},\bm{\chi})\equiv \log \P(\bm{X}_2,...,\X_T|\bm{X}_1,\bm{\nu},\bm{\lambda},\bm{\chi})=  \log\prod_{t=2}^T\P(\X_{t}|\X_{t-1},\bm{\nu},\bm{\lambda},\bm{\chi}) = \\
& = \sum_{t=2}^T\sum_{i=1}^N\log\left(\nu_i\sum_{j=1}^N\lambda_{ij} \delta_{X_t^i,X_{t-1}^j}+(1-\nu_i)(\chi_i)^{X_t^i}(1-\chi_i)^{1-X_t^i}\right),
\end{split}
\ee
by using the chain rule and conditioning on the first observation $\X_1$, where $\delta_{A,B}$ is the Kronecker delta taking value equal to one if $A=B$ and zero otherwise.

MLE of the parameters is obtained by maximizing the log-likelihood via gradient descendent methods \cite{tibi}. Since the number of parameters is of order $O(N^2)$, thus sharply increasing with the number of variables $N$, it is preferable to adopt the solution of the Yule-Walker equations associated with the VDAR(1) model (\ref{vdar1}), instead of a random input, as starting point of the gradient descent method.

\subsubsection{Yule-Walker equations for VDAR(1)}
Similarly to before, The Yule-Walker equations of the VDAR(1) model are formally equivalent to the ones of the VAR(1) model \cite{tsay}
\be\label{yw1vdar1}
\E(\bm{X}_t) = \bm{\phi}_0+\Phi_{1}\E(\bm{X}_{t-1})
\ee
with $\bm{\phi}_0\equiv\{\phi_{0,i}\}_{i=1,...,N}$ and $\Phi_{1}\equiv\{\Phi_{1,ij}\}_{i,j=1,...,N}$, and by using a suitable one-to-one mapping between the two sets of parameters $\{\bm{\nu},\bm{\lambda},\bm{\chi}\}$ and $\{\bm{\phi}_0,\Phi_1\}$. In particular, it is
\be\label{onemapping}
\begin{cases}
\nu_i = \sum_{j=1}^N\Phi_{1,ij},\:\:&\forall i=1,...,N \\
\lambda_{ij} = \frac{\Phi_{1,ij}}{\sum_{j=1}^N\Phi_{1,ij}},\:\:&\forall i,j=1,...,N-1\\
\chi_i = \frac{\phi_{0,i}}{1-\sum_{j=1}^N\Phi_{1,ij}},\:\:&\forall i=1,...,N
\end{cases}
\ee
where $\sum_{j=1}^N\lambda_{ij}=1$, by construction.

It is also trivial to show the following identity
\be\label{ywkvdar1}
\E(\tilde{\X}_t\tilde{\X}_{t-1}') = \Phi_1\E(\tilde{\X}_{t-1}\tilde{\X}_{t-1}'),
\ee
where the tilde is for indicating the mean subtracted variable.

Thus, the Yule-Walker estimator of the parameters $\bm{\phi}_0,\Phi_{1}$ is a standard result of time series analysis, see \cite{tsay} for a reference, and can be obtained by solving the linear system (\ref{ywkvdar1}) for $\Phi_1$, then solving (\ref{yw1vdar1}) for $\bm{\phi}_0$. Finally, the Yule-Walker estimator of the parameters of the V-DAR(1) model is obtained by inverting the one-to-one mapping (\ref{onemapping}).

\subsubsection{Decimation of VDAR(1)}
The MLE $\hat{\bm{\lambda}}$ of the parameter $\{\lambda_{ij}\}_{i,j=1,...,N}$ need to be validated to reveal the causal interactions between the random variables $\{\X^i\}^{i=1,...,N}$, and a possible procedure is the so-called Decimation \cite{decimation,decimationKIM}. It was shown to be working best for the kinetic Ising model \cite{campajola}, which is a logistic regression model for binary random variables. Decimation aims at pruning any parameter which is considered as unnecessary to increase the likelihood. The process starts by setting parameters to zero based on their absolute size, starting with the smallest one, and stops when a transformed likelihood function $\tilde{\ell}$ is maximized,
\begin{equation}
    \tilde{\ell}(q) = \ell (q) - [(1 - q) \ell_{max} + q \ell_0],
\end{equation}

where $q$ is the fraction of pruned parameters, $\ell (q)$ is the log-likelihood (\ref{logPvdar1}) of the model (\ref{vdar1}) with $q$ pruned parameters, $\ell_{max}$ is the log-likelihood of the full model obtained with MLE $\hat{\bm{\lambda}}$, and $\ell_0$ is the log-likelihood (\ref{logPvdar1}) of the VDAR(1) model with parameters $\lambda_{ij}$ set to $0$, $\forall i,j=1,...,N$ (\ie considering data as generated by Bernoulli marginal distributions). The output of the regularization method is a set of validated parameters, in particular a matrix $\tilde{\bm{\lambda}}$ of couplings which is sparser than the initial MLE  $\hat{\bm{\lambda}}$. Thus, a non-zero entry $\tilde{\lambda}_{ij}$ of the matrix $\tilde{\bm{\lambda}}$ describes a Granger causality in tail relation from $j$ to $i$.

\end{document}